\ifx\mnmacrosloaded\undefined 
%
%
%
%

\catcode `\@=11 

\def\@version{1.6}
\def\@verdate{18th September 1995}

%
%


\newif\ifprod@font

\ifx\@typeface\undefined
  \def\@typeface{Comp. Modern}\prod@fontfalse
\else
  \prod@fonttrue 
\fi

\def\newfam{\alloc@8\fam\chardef\sixt@@n} 

\ifprod@font
\font\fiverm=mtr10 at 5pt
\font\fivebf=mtbx10 at 5pt
\font\fiveit=mtti10 at 5pt
\font\fivesl=mtsl10 at 5pt
\font\fivett=cmtt8 at 5pt     \hyphenchar\fivett=-1
\font\fivecsc=mtcsc10 at 5pt
\font\fivesf=mtss10 at 5pt
\font\fivei=mtmi10 at 5pt      \skewchar\fivei='177
\font\fivesy=mtsy10 at 5pt     \skewchar\fivesy='60

\font\sixrm=mtr10 at 6pt
\font\sixbf=mtbx10 at 6pt
\font\sixit=mtti10 at 6pt
\font\sixsl=mtsl10 at 6pt
\font\sixtt=cmtt8 at 6pt      \hyphenchar\sixtt=-1
\font\sixcsc=mtcsc10 at 6pt
\font\sixsf=mtss10 at 6pt
\font\sixi=mtmi10 at 6pt       \skewchar\sixi='177
\font\sixsy=mtsy10 at 6pt      \skewchar\sixsy='60

\font\sevenrm=mtr10 at 7pt
\font\sevenbf=mtbx10 at 7pt
\font\sevenit=mtti10 at 7pt
\font\sevensl=mtsl10 at 7pt
\font\seventt=cmtt8 at 7pt     \hyphenchar\seventt=-1
\font\sevencsc=mtcsc10 at 7pt
\font\sevensf=mtss10 at 7pt
\font\seveni=mtmi10 at 7pt      \skewchar\seveni='177
\font\sevensy=mtsy10 at 7pt     \skewchar\sevensy='60

\font\eightrm=mtr10 at 8pt
\font\eightbf=mtbx10 at 8pt
\font\eightit=mtti10 at 8pt
\font\eighti=mtmi10 at 8pt      \skewchar\eighti='177
\font\eightsy=mtsy10 at 8pt     \skewchar\eightsy='60
\font\eightsl=mtsl10 at 8pt
\font\eighttt=cmtt8             \hyphenchar\eighttt=-1
\font\eightcsc=mtcsc10 at 8pt
\font\eightsf=mtss10 at 8pt

\font\ninerm=mtr10 at 9pt
\font\ninebf=mtbx10 at 9pt
\font\nineit=mtti10 at 9pt
\font\ninei=mtmi10 at 9pt      \skewchar\ninei='177
\font\ninesy=mtsy10 at 9pt     \skewchar\ninesy='60
\font\ninesl=mtsl10 at 9pt
\font\ninett=cmtt9             \hyphenchar\ninett=-1
\font\ninecsc=mtcsc10 at 9pt
\font\ninesf=mtss10 at 9pt

\font\tenrm=mtr10
\font\tenbf=mtbx10
\font\tenit=mtti10
\font\teni=mtmi10		\skewchar\teni='177
\font\tensy=mtsy10		\skewchar\tensy='60
\font\tenex=cmex10
\font\tensl=mtsl10
\font\tentt=cmtt10		\hyphenchar\tentt=-1
\font\tencsc=mtcsc10
\font\tensf=mtss10

\font\elevenrm=mtr10 at 11pt
\font\elevenbf=mtbx10 at 11pt
\font\elevenit=mtti10 at 11pt
\font\eleveni=mtmi10 at 11pt      \skewchar\eleveni='177
\font\elevensy=mtsy10 at 11pt     \skewchar\elevensy='60
\font\elevensl=mtsl10 at 11pt
\font\eleventt=cmtt10 at 11pt     \hyphenchar\eleventt=-1
\font\elevencsc=mtcsc10 at 11pt
\font\elevensf=mtss10 at 11pt

\font\twelverm=mtr10 at 12pt
\font\twelvebf=mtbx10 at 12pt
\font\twelveit=mtti10 at 12pt
\font\twelvesl=mtsl10 at 12pt
\font\twelvett=cmtt12             \hyphenchar\twelvett=-1
\font\twelvecsc=mtcsc10 at 12pt
\font\twelvesf=mtss10 at 12pt
\font\twelvei=mtmi10 at 12pt      \skewchar\twelvei='177
\font\twelvesy=mtsy10 at 12pt     \skewchar\twelvesy='60

\font\fourteenrm=mtr10 at 14pt
\font\fourteenbf=mtbx10 at 14pt
\font\fourteenit=mtti10 at 14pt
\font\fourteeni=mtmi10 at 14pt      \skewchar\fourteeni='177
\font\fourteensy=mtsy10 at 14pt     \skewchar\fourteensy='60
\font\fourteensl=mtsl10 at 14pt
\font\fourteentt=cmtt12 at 14pt     \hyphenchar\fourteentt=-1
\font\fourteencsc=mtcsc10 at 14pt
\font\fourteensf=mtss10 at 14pt

\font\seventeenrm=mtr10 at 17pt
\font\seventeenbf=mtbx10 at 17pt
\font\seventeenit=mtti10 at 17pt
\font\seventeeni=mtmi10 at 17pt      \skewchar\seventeeni='177
\font\seventeensy=mtsy10 at 17pt     \skewchar\seventeensy='60
\font\seventeensl=mtsl10 at 17pt
\font\seventeentt=cmtt12 at 17pt     \hyphenchar\seventeentt=-1
\font\seventeencsc=mtcsc10 at 17pt
\font\seventeensf=mtss10 at 17pt
\else
\font\fiverm=cmr5
\font\fivei=cmmi5             \skewchar\fivei='177
\font\fivesy=cmsy5            \skewchar\fivesy='60
\font\fivebf=cmbx5

\font\sixrm=cmr6
\font\sixi=cmmi6             \skewchar\sixi='177
\font\sixsy=cmsy6            \skewchar\sixsy='60
\font\sixbf=cmbx6

\font\sevenrm=cmr7
\font\sevenit=cmti7
\font\seveni=cmmi7             \skewchar\seveni='177
\font\sevensy=cmsy7            \skewchar\sevensy='60
\font\sevenbf=cmbx7

\font\eightrm=cmr8
\font\eightbf=cmbx8
\font\eightit=cmti8
\font\eighti=cmmi8			\skewchar\eighti='177
\font\eightsy=cmsy8			\skewchar\eightsy='60
\font\eightsl=cmsl8
\font\eighttt=cmtt8			\hyphenchar\eighttt=-1
\font\eightcsc=cmcsc10 at 8pt
\font\eightsf=cmss8

\font\ninerm=cmr9
\font\ninebf=cmbx9
\font\nineit=cmti9
\font\ninei=cmmi9			\skewchar\ninei='177
\font\ninesy=cmsy9			\skewchar\ninesy='60
\font\ninesl=cmsl9
\font\ninett=cmtt9			\hyphenchar\ninett=-1
\font\ninecsc=cmcsc10 at 9pt
\font\ninesf=cmss9

\font\tenrm=cmr10
\font\tenbf=cmbx10
\font\tenit=cmti10
\font\teni=cmmi10		\skewchar\teni='177
\font\tensy=cmsy10		\skewchar\tensy='60
\font\tenex=cmex10
\font\tensl=cmsl10
\font\tentt=cmtt10		\hyphenchar\tentt=-1
\font\tencsc=cmcsc10
\font\tensf=cmss10

\font\elevenrm=cmr10 scaled \magstephalf
\font\elevenbf=cmbx10 scaled \magstephalf
\font\elevenit=cmti10 scaled \magstephalf
\font\eleveni=cmmi10 scaled \magstephalf	\skewchar\eleveni='177
\font\elevensy=cmsy10 scaled \magstephalf	\skewchar\elevensy='60
\font\elevensl=cmsl10 scaled \magstephalf
\font\eleventt=cmtt10 scaled \magstephalf	\hyphenchar\eleventt=-1
\font\elevencsc=cmcsc10 scaled \magstephalf
\font\elevensf=cmss10 scaled \magstephalf

\font\twelverm=cmr10 scaled \magstep1
\font\twelvebf=cmbx10 scaled \magstep1
\font\twelvei=cmmi10 scaled \magstep1      \skewchar\twelvei='177
\font\twelvesy=cmsy10 scaled \magstep1     \skewchar\twelvesy='60

\font\fourteenrm=cmr10 scaled \magstep2
\font\fourteenbf=cmbx10 scaled \magstep2
\font\fourteenit=cmti10 scaled \magstep2
\font\fourteeni=cmmi10 scaled \magstep2		\skewchar\fourteeni='177
\font\fourteensy=cmsy10 scaled \magstep2	\skewchar\fourteensy='60
\font\fourteensl=cmsl10 scaled \magstep2
\font\fourteentt=cmtt10 scaled \magstep2	\hyphenchar\fourteentt=-1
\font\fourteencsc=cmcsc10 scaled \magstep2
\font\fourteensf=cmss10 scaled \magstep2

\font\seventeenrm=cmr10 scaled \magstep3
\font\seventeenbf=cmbx10 scaled \magstep3
\font\seventeenit=cmti10 scaled \magstep3
\font\seventeeni=cmmi10 scaled \magstep3	\skewchar\seventeeni='177
\font\seventeensy=cmsy10 scaled \magstep3	\skewchar\seventeensy='60
\font\seventeensl=cmsl10 scaled \magstep3
\font\seventeentt=cmtt10 scaled \magstep3	\hyphenchar\seventeentt=-1
\font\seventeencsc=cmcsc10 scaled \magstep3
\font\seventeensf=cmss10 scaled \magstep3
\fi

\def\hexnumber#1{\ifcase#1 0\or1\or2\or3\or4\or5\or6\or7\or8\or9\or
  A\or B\or C\or D\or E\or F\fi}

\def\makestrut{%
  \setbox\strutbox=\hbox{%
    \vrule height.7\baselineskip depth.3\baselineskip width \z@}%
}

\def\baselinestretch{1}
\newskip\tmp@bls

\def\b@ls#1{
  \tmp@bls=#1\relax
  \baselineskip=#1\relax\makestrut
  \normalbaselineskip=\baselinestretch\tmp@bls
  \normalbaselines
}

\def\nostb@ls#1{
  \normalbaselineskip=#1\relax
  \normalbaselines
  \makestrut
}

%

\newfam\scfam  
\newfam\sffam  

\def\mit{\fam\@ne}
\def\cal{\fam\tw@}
\def\em{\ifdim\fontdimen1\font>\z@ \rm\else\it\fi}

\textfont3=\tenex
\scriptfont3=\tenex
\scriptscriptfont3=\tenex

\setbox0=\hbox{\tenex B} \p@renwd=\wd0 

\def\eightpoint{
  \def\rm{\fam0\eightrm}%
  \textfont0=\eightrm \scriptfont0=\sixrm \scriptscriptfont0=\fiverm%
  \textfont1=\eighti  \scriptfont1=\sixi  \scriptscriptfont1=\fivei%
  \textfont2=\eightsy \scriptfont2=\sixsy \scriptscriptfont2=\fivesy%
  \textfont\itfam=\eightit\def\it{\fam\itfam\eightit}%
  \ifprod@font
    \scriptfont\itfam=\sixit
      \scriptscriptfont\itfam=\fiveit
  \else
    \scriptfont\itfam=\eightit
      \scriptscriptfont\itfam=\eightit
  \fi
  \textfont\bffam=\eightbf%
    \scriptfont\bffam=\sixbf%
      \scriptscriptfont\bffam=\fivebf%
  \def\bf{\fam\bffam\eightbf}%
  \textfont\slfam=\eightsl\def\sl{\fam\slfam\eightsl}%
  \ifprod@font
    \scriptfont\slfam=\sixsl
      \scriptscriptfont\slfam=\fivesl
  \else
    \scriptfont\slfam=\eightsl
      \scriptscriptfont\slfam=\eightsl
  \fi
  \textfont\ttfam=\eighttt\def\tt{\fam\ttfam\eighttt}%
  \ifprod@font
    \scriptfont\ttfam=\sixtt
      \scriptscriptfont\ttfam=\fivett
  \else
    \scriptfont\ttfam=\eighttt
      \scriptscriptfont\ttfam=\eighttt
  \fi
  \textfont\scfam=\eightcsc\def\sc{\fam\scfam\eightcsc}%
  \ifprod@font
    \scriptfont\scfam=\sixcsc
      \scriptscriptfont\scfam=\fivecsc
  \else
    \scriptfont\scfam=\eightcsc
      \scriptscriptfont\scfam=\eightcsc
  \fi
  \textfont\sffam=\eightsf\def\sf{\fam\sffam\eightsf}%
  \ifprod@font
    \scriptfont\sffam=\sixsf
      \scriptscriptfont\sffam=\fivesf
  \else
    \scriptfont\sffam=\eightsf
      \scriptscriptfont\sffam=\eightsf
  \fi
  \def\oldstyle{\fam\@ne\eighti}%
  \b@ls{10pt}\rm\@viiipt%
}
\def\@viiipt{}

\def\ninepoint{
  \def\rm{\fam0\ninerm}%
  \textfont0=\ninerm \scriptfont0=\sixrm \scriptscriptfont0=\fiverm%
  \textfont1=\ninei  \scriptfont1=\sixi  \scriptscriptfont1=\fivei%
  \textfont2=\ninesy \scriptfont2=\sixsy \scriptscriptfont2=\fivesy%
  \textfont\itfam=\nineit\def\it{\fam\itfam\nineit}%
  \ifprod@font
    \scriptfont\itfam=\sixit
      \scriptscriptfont\itfam=\fiveit
  \else
    \scriptfont\itfam=\nineit
      \scriptscriptfont\itfam=\nineit
  \fi
  \textfont\bffam=\ninebf%
    \scriptfont\bffam=\sixbf%
      \scriptscriptfont\bffam=\fivebf%
  \def\bf{\fam\bffam\ninebf}%
  \textfont\slfam=\ninesl\def\sl{\fam\slfam\ninesl}%
  \ifprod@font
    \scriptfont\slfam=\sixsl
      \scriptscriptfont\slfam=\fivesl
  \else
    \scriptfont\slfam=\ninesl
      \scriptscriptfont\slfam=\ninesl
  \fi
  \textfont\ttfam=\ninett\def\tt{\fam\ttfam\ninett}%
  \ifprod@font
    \scriptfont\ttfam=\sixtt
      \scriptscriptfont\ttfam=\fivett
  \else
    \scriptfont\ttfam=\ninett
      \scriptscriptfont\ttfam=\ninett
  \fi
  \textfont\scfam=\ninecsc\def\sc{\fam\scfam\ninecsc}%
  \ifprod@font
    \scriptfont\scfam=\sixcsc
      \scriptscriptfont\scfam=\fivecsc
  \else
    \scriptfont\scfam=\ninecsc
      \scriptscriptfont\scfam=\ninecsc
  \fi
  \textfont\sffam=\ninesf\def\sf{\fam\sffam\ninesf}%
  \ifprod@font
    \scriptfont\sffam=\sixsf
      \scriptscriptfont\sffam=\fivesf
  \else
    \scriptfont\sffam=\ninesf
      \scriptscriptfont\sffam=\ninesf
  \fi
  \def\oldstyle{\fam\@ne\ninei}%
  \b@ls{\TextLeading plus \Feathering}\rm\@ixpt%
}
\def\@ixpt{}

\def\tenpoint{
  \def\rm{\fam0\tenrm}%
  \textfont0=\tenrm \scriptfont0=\sevenrm \scriptscriptfont0=\fiverm%
  \textfont1=\teni  \scriptfont1=\seveni  \scriptscriptfont1=\fivei%
  \textfont2=\tensy \scriptfont2=\sevensy \scriptscriptfont2=\fivesy%
  \textfont\itfam=\tenit\def\it{\fam\itfam\tenit}%
  \ifprod@font
    \scriptfont\itfam=\sevenit
      \scriptscriptfont\itfam=\fiveit
  \else
    \scriptfont\itfam=\tenit
      \scriptscriptfont\itfam=\tenit
  \fi
  \textfont\bffam=\tenbf%
    \scriptfont\bffam=\sevenbf%
      \scriptscriptfont\bffam=\fivebf%
  \def\bf{\fam\bffam\tenbf}%
  \textfont\slfam=\tensl\def\sl{\fam\slfam\tensl}%
  \ifprod@font
    \scriptfont\slfam=\sevensl
      \scriptscriptfont\slfam=\fivesl
  \else
    \scriptfont\slfam=\tensl
      \scriptscriptfont\slfam=\tensl
  \fi
  \textfont\ttfam=\tentt\def\tt{\fam\ttfam\tentt}%
  \ifprod@font
    \scriptfont\ttfam=\seventt
      \scriptscriptfont\ttfam=\fivett
  \else
    \scriptfont\ttfam=\tentt
      \scriptscriptfont\ttfam=\tentt
  \fi
  \textfont\scfam=\tencsc\def\sc{\fam\scfam\tencsc}%
  \ifprod@font
    \scriptfont\scfam=\sevencsc
      \scriptscriptfont\scfam=\fivecsc
  \else
    \scriptfont\scfam=\tencsc
      \scriptscriptfont\scfam=\tencsc
  \fi
  \textfont\sffam=\tensf\def\sf{\fam\sffam\tensf}%
  \ifprod@font
    \scriptfont\sffam=\sevensf
      \scriptscriptfont\sffam=\fivesf
  \else
    \scriptfont\sffam=\tensf
      \scriptscriptfont\sffam=\tensf
  \fi
  \def\oldstyle{\fam\@ne\teni}%
  \b@ls{11pt}\rm\@xpt%
}
\def\@xpt{}

\def\elevenpoint{
  \def\rm{\fam0\elevenrm}%
  \textfont0=\elevenrm \scriptfont0=\eightrm \scriptscriptfont0=\sixrm%
  \textfont1=\eleveni  \scriptfont1=\eighti  \scriptscriptfont1=\sixi%
  \textfont2=\elevensy \scriptfont2=\eightsy \scriptscriptfont2=\sixsy%
  \textfont\itfam=\elevenit\def\it{\fam\itfam\elevenit}%
  \ifprod@font
    \scriptfont\itfam=\eightit
      \scriptscriptfont\itfam=\sixit
  \else
    \scriptfont\itfam=\elevenit
      \scriptscriptfont\itfam=\elevenit
  \fi
  \textfont\bffam=\elevenbf%
    \scriptfont\bffam=\eightbf%
      \scriptscriptfont\bffam=\sixbf%
  \def\bf{\fam\bffam\elevenbf}%
  \textfont\slfam=\elevensl\def\sl{\fam\slfam\elevensl}%
  \ifprod@font
    \scriptfont\slfam=\eightsl
      \scriptscriptfont\slfam=\sixsl
  \else
    \scriptfont\slfam=\elevensl
      \scriptscriptfont\slfam=\elevensl
  \fi
  \textfont\ttfam=\eleventt\def\tt{\fam\ttfam\eleventt}%
  \ifprod@font
    \scriptfont\ttfam=\eighttt
      \scriptscriptfont\ttfam=\sixtt
  \else
    \scriptfont\ttfam=\eleventt
      \scriptscriptfont\ttfam=\eleventt
  \fi
  \textfont\scfam=\elevencsc\def\sc{\fam\scfam\elevencsc}%
  \ifprod@font
    \scriptfont\scfam=\eightcsc
      \scriptscriptfont\scfam=\sixcsc
  \else
    \scriptfont\scfam=\elevencsc
      \scriptscriptfont\scfam=\elevencsc
  \fi
  \textfont\sffam=\elevensf\def\sf{\fam\sffam\elevensf}%
  \ifprod@font
    \scriptfont\sffam=\eightsf
      \scriptscriptfont\sffam=\sixsf
  \else
    \scriptfont\sffam=\elevensf
      \scriptscriptfont\sffam=\elevensf
  \fi
  \def\oldstyle{\fam\@ne\eleveni}%
  \b@ls{13pt}\rm\@xipt%
}
\def\@xipt{}

\def\fourteenpoint{
  \def\rm{\fam0\fourteenrm}%
  \textfont0\fourteenrm  \scriptfont0\tenrm  \scriptscriptfont0\sevenrm%
  \textfont1\fourteeni   \scriptfont1\teni   \scriptscriptfont1\seveni%
  \textfont2\fourteensy  \scriptfont2\tensy  \scriptscriptfont2\sevensy%
  \textfont\itfam=\fourteenit\def\it{\fam\itfam\fourteenit}%
  \ifprod@font
    \scriptfont\itfam=\tenit
      \scriptscriptfont\itfam=\sevenit
  \else
    \scriptfont\itfam=\fourteenit
      \scriptscriptfont\itfam=\fourteenit
  \fi
  \textfont\bffam=\fourteenbf%
    \scriptfont\bffam=\tenbf%
      \scriptscriptfont\bffam=\sevenbf%
  \def\bf{\fam\bffam\fourteenbf}%
  \textfont\slfam=\fourteensl\def\sl{\fam\slfam\fourteensl}%
  \ifprod@font
    \scriptfont\slfam=\tensl
      \scriptscriptfont\slfam=\sevensl
  \else
    \scriptfont\slfam=\fourteensl
      \scriptscriptfont\slfam=\fourteensl
  \fi
  \textfont\ttfam=\fourteentt\def\tt{\fam\ttfam\fourteentt}%
  \ifprod@font
    \scriptfont\ttfam=\tentt
      \scriptscriptfont\ttfam=\seventt
  \else
    \scriptfont\ttfam=\fourteentt
      \scriptscriptfont\ttfam=\fourteentt
  \fi
  \textfont\scfam=\fourteencsc\def\sc{\fam\scfam\fourteencsc}%
  \ifprod@font
    \scriptfont\scfam=\tencsc
      \scriptscriptfont\scfam=\sevencsc
  \else
    \scriptfont\scfam=\fourteencsc
      \scriptscriptfont\scfam=\fourteencsc
  \fi
  \textfont\sffam=\fourteensf\def\sf{\fam\sffam\fourteensf}%
  \ifprod@font
    \scriptfont\sffam=\tensf
      \scriptscriptfont\sffam=\sevensf
  \else
    \scriptfont\sffam=\fourteensf
      \scriptscriptfont\sffam=\fourteensf
  \fi
  \def\oldstyle{\fam\@ne\fourteeni}%
  \b@ls{17pt}\rm\@xivpt%
}
\def\@xivpt{}

\def\seventeenpoint{
  \def\rm{\fam0\seventeenrm}%
  \textfont0\seventeenrm  \scriptfont0\twelverm  \scriptscriptfont0\tenrm%
  \textfont1\seventeeni   \scriptfont1\twelvei   \scriptscriptfont1\teni%
  \textfont2\seventeensy  \scriptfont2\twelvesy  \scriptscriptfont2\tensy%
  \textfont\itfam=\seventeenit\def\it{\fam\itfam\seventeenit}%
  \ifprod@font
    \scriptfont\itfam=\twelveit
      \scriptscriptfont\itfam=\tenit
  \else
    \scriptfont\itfam=\seventeenit
      \scriptscriptfont\itfam=\seventeenit
  \fi
  \textfont\bffam=\seventeenbf%
    \scriptfont\bffam=\twelvebf%
      \scriptscriptfont\bffam=\tenbf%
  \def\bf{\fam\bffam\seventeenbf}%
  \textfont\slfam=\seventeensl\def\sl{\fam\slfam\seventeensl}%
  \ifprod@font
    \scriptfont\slfam=\twelvesl
      \scriptscriptfont\slfam=\tensl
  \else
    \scriptfont\slfam=\seventeensl
      \scriptscriptfont\slfam=\seventeensl
  \fi
  \textfont\ttfam=\seventeentt\def\tt{\fam\ttfam\seventeentt}%
  \ifprod@font
    \scriptfont\ttfam=\twelvett
      \scriptscriptfont\ttfam=\tentt
  \else
    \scriptfont\ttfam=\seventeentt
      \scriptscriptfont\ttfam=\seventeentt
  \fi
  \textfont\scfam=\seventeencsc\def\sc{\fam\scfam\seventeencsc}%
  \ifprod@font
    \scriptfont\scfam=\twelvecsc
      \scriptscriptfont\scfam=\tencsc
  \else
    \scriptfont\scfam=\seventeencsc
      \scriptscriptfont\scfam=\seventeencsc
  \fi
  \textfont\sffam=\seventeensf\def\sf{\fam\sffam\seventeensf}%
  \ifprod@font
    \scriptfont\sffam=\twelvesf
      \scriptscriptfont\sffam=\tensf
  \else
    \scriptfont\sffam=\seventeensf
      \scriptscriptfont\sffam=\seventeensf
  \fi
  \def\oldstyle{\fam\@ne\seventeeni}%
  \b@ls{20pt}\rm\@xviipt%
}
\def\@xviipt{}

\lineskip=1pt      \normallineskip=\lineskip
\lineskiplimit=\z@ \normallineskiplimit=\lineskiplimit


\def\loadboldmathnames{%
  \def\balpha{{\bmath{\alpha}}}%
  \def\bbeta{{\bmath{\beta}}}%
  \def\bgamma{{\bmath{\gamma}}}%
  \def\bdelta{{\bmath{\delta}}}%
  \def\bepsilon{{\bmath{\epsilon}}}%
  \def\bzeta{{\bmath{\zeta}}}%
  \def\boldeta{{\bmath{\eta}}}%
  \def\btheta{{\bmath{\theta}}}%
  \def\biota{{\bmath{\iota}}}%
  \def\bkappa{{\bmath{\kappa}}}%
  \def\blambda{{\bmath{\lambda}}}%
  \def\bmu{{\bmath{\mu}}}%
  \def\bnu{{\bmath{\nu}}}%
  \def\bxi{{\bmath{\xi}}}%
  \def\bpi{{\bmath{\pi}}}%
  \def\brho{{\bmath{\rho}}}%
  \def\bsigma{{\bmath{\sigma}}}%
  \def\btau{{\bmath{\tau}}}%
  \def\bupsilon{{\bmath{\upsilon}}}%
  \def\bphi{{\bmath{\phi}}}%
  \def\bchi{{\bmath{\chi}}}%
  \def\bpsi{{\bmath{\psi}}}%
  \def\bomega{{\bmath{\omega}}}%
  \def\bvarepsilon{{\bmath{\varepsilon}}}%
  \def\bvartheta{{\bmath{\vartheta}}}%
  \def\bvarpi{{\bmath{\varpi}}}%
  \def\bvarrho{{\bmath{\varrho}}}%
  \def\bvarsigma{{\bmath{\varsigma}}}%
  \def\bvarphi{{\bmath{\varphi}}}%
  \def\baleph{{\bmath{\aleph}}}%
  \def\bimath{{\bmath{\imath}}}%
  \def\bjmath{{\bmath{\jmath}}}%
  \def\bell{{\bmath{\ell}}}%
  \def\bwp{{\bmath{\wp}}}%
  \def\bRe{{\bmath{\Re}}}%
  \def\bIm{{\bmath{\Im}}}%
  \def\bpartial{{\bmath{\partial}}}%
  \def\binfty{{\bmath{\infty}}}%
  \def\bprime{{\bmath{\prime}}}%
  \def\bemptyset{{\bmath{\emptyset}}}%
  \def\bnabla{{\bmath{\nabla}}}%
  \def\btop{{\bmath{\top}}}%
  \def\bbot{{\bmath{\bot}}}%
  \def\btriangle{{\bmath{\triangle}}}%
  \def\bforall{{\bmath{\forall}}}%
  \def\bexists{{\bmath{\exists}}}%
  \def\bneg{{\bmath{\neg}}}%
  \def\bflat{{\bmath{\flat}}}%
  \def\bnatural{{\bmath{\natural}}}%
  \def\bsharp{{\bmath{\sharp}}}%
  \def\bclubsuit{{\bmath{\clubsuit}}}%
  \def\bdiamondsuit{{\bmath{\diamondsuit}}}%
  \def\bheartsuit{{\bmath{\heartsuit}}}%
  \def\bspadesuit{{\bmath{\spadesuit}}}%
  \def\bsmallint{{\bmath{\smallint}}}%
  \def\btriangleleft{{\bmath{\triangleleft}}}%
  \def\btriangleright{{\bmath{\triangleright}}}%
  \def\bbigtriangleup{{\bmath{\bigtriangleup}}}%
  \def\bbigtriangledown{{\bmath{\bigtriangledown}}}%
  \def\bwedge{{\bmath{\wedge}}}%
  \def\bvee{{\bmath{\vee}}}%
  \def\bcap{{\bmath{\cap}}}%
  \def\bcup{{\bmath{\cup}}}%
  \def\bddagger{{\bmath{\ddagger}}}%
  \def\bdagger{{\bmath{\dagger}}}%
  \def\bsqcap{{\bmath{\sqcap}}}%
  \def\bsqcup{{\bmath{\sqcup}}}%
  \def\buplus{{\bmath{\uplus}}}%
  \def\bamalg{{\bmath{\amalg}}}%
  \def\bdiamond{{\bmath{\diamond}}}%
  \def\bbullet{{\bmath{\bullet}}}%
  \def\bwr{{\bmath{\wr}}}%
  \def\bdiv{{\bmath{\div}}}%
  \def\bodot{{\bmath{\odot}}}%
  \def\boslash{{\bmath{\oslash}}}%
  \def\botimes{{\bmath{\otimes}}}%
  \def\bominus{{\bmath{\ominus}}}%
  \def\boplus{{\bmath{\oplus}}}%
  \def\bmp{{\bmath{\mp}}}%
  \def\bpm{{\bmath{\pm}}}%
  \def\bcirc{{\bmath{\circ}}}%
  \def\bbigcirc{{\bmath{\bigcirc}}}%
  \def\bsetminus{{\bmath{\setminus}}}%
  \def\bcdot{{\bmath{\cdot}}}%
  \def\bast{{\bmath{\ast}}}%
  \def\btimes{{\bmath{\times}}}%
  \def\bstar{{\bmath{\star}}}%
  \def\bpropto{{\bmath{\propto}}}%
  \def\bsqsubseteq{{\bmath{\sqsubseteq}}}%
  \def\bsqsupseteq{{\bmath{\sqsupseteq}}}%
  \def\bparallel{{\bmath{\parallel}}}%
  \def\bmid{{\bmath{\mid}}}%
  \def\bdashv{{\bmath{\dashv}}}%
  \def\bvdash{{\bmath{\vdash}}}%
  \def\bnearrow{{\bmath{\nearrow}}}%
  \def\bsearrow{{\bmath{\searrow}}}%
  \def\bnwarrow{{\bmath{\nwarrow}}}%
  \def\bswarrow{{\bmath{\swarrow}}}%
  \def\bLeftrightarrow{{\bmath{\Leftrightarrow}}}%
  \def\bLeftarrow{{\bmath{\Leftarrow}}}%
  \def\bRightarrow{{\bmath{\Rightarrow}}}%
  \def\bleq{{\bmath{\leq}}}%
  \def\bgeq{{\bmath{\geq}}}%
  \def\bsucc{{\bmath{\succ}}}%
  \def\bprec{{\bmath{\prec}}}%
  \def\bapprox{{\bmath{\approx}}}%
  \def\bsucceq{{\bmath{\succeq}}}%
  \def\bpreceq{{\bmath{\preceq}}}%
  \def\bsupset{{\bmath{\supset}}}%
  \def\bsubset{{\bmath{\subset}}}%
  \def\bsupseteq{{\bmath{\supseteq}}}%
  \def\bsubseteq{{\bmath{\subseteq}}}%
  \def\bin{{\bmath{\in}}}%
  \def\bni{{\bmath{\ni}}}%
  \def\bgg{{\bmath{\gg}}}%
  \def\bll{{\bmath{\ll}}}%
  \def\bnot{{\bmath{\not}}}%
  \def\bleftrightarrow{{\bmath{\leftrightarrow}}}%
  \def\bleftarrow{{\bmath{\leftarrow}}}%
  \def\brightarrow{{\bmath{\rightarrow}}}%
  \def\bmapstochar{{\bmath{\mapstochar}}}%
  \def\bsim{{\bmath{\sim}}}%
  \def\bsimeq{{\bmath{\simeq}}}%
  \def\bperp{{\bmath{\perp}}}%
  \def\bequiv{{\bmath{\equiv}}}%
  \def\basymp{{\bmath{\asymp}}}%
  \def\bsmile{{\bmath{\smile}}}%
  \def\bfrown{{\bmath{\frown}}}%
  \def\bleftharpoonup{{\bmath{\leftharpoonup}}}%
  \def\bleftharpoondown{{\bmath{\leftharpoondown}}}%
  \def\brightharpoonup{{\bmath{\rightharpoonup}}}%
  \def\brightharpoondown{{\bmath{\rightharpoondown}}}%
  \def\blhook{{\bmath{\lhook}}}%
  \def\brhook{{\bmath{\rhook}}}%
  \def\bldotp{{\bmath{\ldotp}}}%
  \def\bcdotp{{\bmath{\cdotp}}}%
}

\def\,{\relax\ifmmode \mskip\thinmuskip\else \thinspace\fi}
\let\protect=\relax

\long\def\@ifundefined#1#2#3{\expandafter\ifx\csname
  #1\endcsname\relax#2\else#3\fi}




\newtoks\math@groups \math@groups={}
\def\addtom@thgroup#1#2{#1\expandafter{\the#1#2}} 



\def\addtosizeh@ok#1#2#3#4{%
  \expandafter\def\csname @#1pt\endcsname{%
    \def\s@ze{#2}\def\ss@ze{#3}\def\sss@ze{#4}\the\math@groups%
  }%
}



\let\resetsizehook=\addtosizeh@ok


\ifprod@font
  \addtosizeh@ok{viii} {8} {6}  {5}
  \addtosizeh@ok{ix}   {9} {6}  {5}
  \addtosizeh@ok{x}    {10}{7}  {5}
  \addtosizeh@ok{xi}   {11}{8}  {6}
  \addtosizeh@ok{xiv}  {14}{10} {7}
  \addtosizeh@ok{xvii} {17}{12}{10}
\else
  \addtosizeh@ok{viii} {8}     {6}     {5}
  \addtosizeh@ok{ix}   {9}     {6}     {5}
  \addtosizeh@ok{x}    {10}    {7}     {5}
  \addtosizeh@ok{xi}   {10.95} {8}     {6}
  \addtosizeh@ok{xiv}  {14.4}  {10}    {7}
  \addtosizeh@ok{xvii} {17.28} {12}    {10}
\fi

\def\get@font#1#2#3{%
  \edef\fonts@ze{\romannumeral#3}
  \edef\fontn@me{\fonts@ze#1}
  \@ifundefined{\fontn@me}%
    {
     \global\expandafter\font\csname \fontn@me\endcsname=#2 at #3pt}%
    {}%
}

\def\ass@tfont#1#2{%
  \xdef\fam@name{\csname #1\endcsname}%
  \xdef\font@name{\csname #2\endcsname}%
  \let\textfont@name\font@name
  \textfont\fam@name\textfont@name
}

\def\ass@sfont#1#2{%
  \xdef\fam@name{\csname #1\endcsname}%
  \xdef\font@name{\csname #2\endcsname}%
  \let\textfont@name\font@name
  \scriptfont\fam@name\textfont@name
}

\def\ass@ssfont#1#2{%
  \xdef\fam@name{\csname #1\endcsname}%
  \xdef\font@name{\csname #2\endcsname}%
  \let\textfont@name\font@name
  \scriptscriptfont\fam@name\textfont@name
}


\def\NewSymbolFont#1#2{%
  \expandafter\ifx\csname sym#1fam\endcsname\relax 
    \expandafter\newfam\csname sym#1fam\endcsname
    \expandafter\edef\csname sym#1fam\endcsname{\the\allocationnumber}%
    \addtom@thgroup\math@groups{%
      \get@font{#1}{#2}{\s@ze}%
      \ass@tfont{sym#1fam}{\fontn@me}%
      \get@font{#1}{#2}{\ss@ze}%
      \ass@sfont{sym#1fam}{\fontn@me}%
      \get@font{#1}{#2}{\sss@ze}%
      \ass@ssfont{sym#1fam}{\fontn@me}%
    }%
  \else
    \errmessage{Family `#1' already defined}%
  \fi
}


\def\NewMathSymbol#1#2#3#4{%
  \edef\f@mly{\expandafter\hexnumber{\csname sym#3fam\endcsname}}%
  \mathchardef#1="#2\f@mly#4\relax
}


\newif\ifd@f

\def\NewMathDelimiter#1#2#3#4#5#6{%
  \d@ftrue
  \expandafter\ifx\csname sym#3fam\endcsname\relax
    \d@ffalse \errmessage{Family `#3' is not defined}%
  \fi
  \expandafter\ifx\csname sym#5fam\endcsname\relax
    \d@ffalse \errmessage{Family `#5' is not defined}%
  \fi
  \ifd@f
    \edef\f@mly{\expandafter\hexnumber{\csname sym#3fam\endcsname}}%
    \edef\f@mlytw@{\expandafter\hexnumber{\csname sym#5fam\endcsname}}%
    \xdef#1{\delimiter"#2\f@mly #4\f@mlytw@ #6\relax}%
  \fi
}


\def\setboxz@h{\setbox\z@\hbox}
\def\wdz@{\wd\z@}
\def\boxz@{\box\z@}
\def\setbox@ne{\setbox\@ne}
\def\wd@ne{\wd\@ne}

\def\math@atom#1#2{%
   \binrel@{#1}\binrel@@{#2}}
\def\binrel@#1{\setboxz@h{\thinmuskip0mu
  \medmuskip\m@ne mu\thickmuskip\@ne mu$#1\m@th$}%
 \setbox@ne\hbox{\thinmuskip0mu\medmuskip\m@ne mu\thickmuskip
  \@ne mu${}#1{}\m@th$}%
 \setbox\tw@\hbox{\hskip\wd@ne\hskip-\wdz@}}
\def\binrel@@#1{\ifdim\wd2<\z@\mathbin{#1}\else\ifdim\wd\tw@>\z@
 \mathrel{#1}\else{#1}\fi\fi}

\def\m@thit{1}

\def\set@skchar#1{\global\expandafter\skewchar
  \csname\fontn@me\endcsname=#1\relax}

\def\NewMathAlphabet#1#2#3{%
  \def\tst{#3}%
  \ifx\tst\empty\else 
    \expandafter\gdef\csname #1@sc\endcsname{}
  \fi
  \expandafter\def\csname #1\endcsname{
    \protect\csname @#1\endcsname}%
  \expandafter\def\csname @#1\endcsname##1{
    {%
    \begingroup
      \get@font{#1}{#2}{\s@ze}%
      \@ifundefined{#1@sc}{}{\set@skchar{#3}}%
      \ass@tfont{m@thit}{\fontn@me}%
      \get@font{#1}{#2}{\ss@ze}%
      \@ifundefined{#1@sc}{}{\set@skchar{#3}}%
      \ass@sfont{m@thit}{\fontn@me}%
      \get@font{#1}{#2}{\sss@ze}%
      \@ifundefined{#1@sc}{}{\set@skchar{#3}}%
      \ass@ssfont{m@thit}{\fontn@me}%
      \math@atom{##1}{%
      \mathchoice%
        {\hbox{$\m@th\displaystyle##1$}}%
        {\hbox{$\m@th\textstyle##1$}}%
        {\hbox{$\m@th\scriptstyle##1$}}%
        {\hbox{$\m@th\scriptscriptstyle##1$}}}%
    \endgroup
    }%
  }%
}


\newif\iffirstta  \firsttatrue

\def\set@hchar#1{\global\expandafter\hyphenchar
  \csname\fontn@me\endcsname=#1\relax}

\def\NewTextAlphabet#1#2#3{%
  \iffirstta
    \global\firsttafalse
    \newfam\scratchfam
    \edef\scrt@fam{\the\allocationnumber}%
  \fi
  \def\tst{#3}%
  \ifx\tst\empty\else 
    \expandafter\gdef\csname #1@hc\endcsname{}
  \fi
  \expandafter\def\csname #1\endcsname{
    \protect\csname t@#1\endcsname}%
  \long\expandafter\def\csname t@#1\endcsname##1{
    \ifmmode
      \typeout{Warning: do not use \expandafter\string\csname #1\endcsname
        \space in math mode}\fi%
    {%
      \get@font{#1}{#2}{\s@ze}\let\t@xtfnt=\fontn@me\relax
      \@ifundefined{#1@hc}{}{\set@hchar{#3}}%
      \ass@tfont{scrt@fam}{\fontn@me}%
      \get@font{#1}{#2}{\ss@ze}%
      \@ifundefined{#1@hc}{}{\set@hchar{#3}}%
      \ass@sfont{scrt@fam}{\fontn@me}%
      \get@font{#1}{#2}{\sss@ze}%
      \@ifundefined{#1@hc}{}{\set@hchar{#3}}%
      \ass@ssfont{scrt@fam}{\fontn@me}%
      \fam\scratchfam\csname\t@xtfnt\endcsname
    ##1%
    }%
  }%
  \expandafter\def\csname #1shape
    \endcsname{\protect\csname @#1shape\endcsname}%
  \expandafter\def\csname @#1shape\endcsname{
    \ifmmode
      \typeout{Warning: do not use \expandafter\string\csname
        #1shape\endcsname \space in math mode}\fi
      \get@font{#1}{#2}{\s@ze}\let\t@xtfnt=\fontn@me\relax
      \@ifundefined{#1@hc}{}{\set@hchar{#3}}%
      \ass@tfont{scrt@fam}{\fontn@me}%
      \get@font{#1}{#2}{\ss@ze}%
      \@ifundefined{#1@hc}{}{\set@hchar{#3}}%
      \ass@sfont{scrt@fam}{\fontn@me}%
      \get@font{#1}{#2}{\sss@ze}%
      \@ifundefined{#1@hc}{}{\set@hchar{#3}}%
      \ass@ssfont{scrt@fam}{\fontn@me}%
      \fam\scratchfam\csname\t@xtfnt\endcsname
  }%
}


\ifprod@font
  \def\math@itfnt{mtmib10}
  \def\math@syfnt{mtbsy10}
\else
  \def\math@itfnt{cmmib10}
  \def\math@syfnt{cmbsy10}
\fi

\def\m@thsy{2}

\def\bmath{\protect\@bmath}
\def\@bmath#1{%
  {%
  \begingroup
    \get@font{mthit}{\math@itfnt}{\s@ze}\set@skchar{'177}%
    \ass@tfont{m@thit}{\fontn@me}%
    \get@font{mthit}{\math@itfnt}{\ss@ze}\set@skchar{'177}%
    \ass@sfont{m@thit}{\fontn@me}%
    \get@font{mthit}{\math@itfnt}{\sss@ze}\set@skchar{'177}%
    \ass@ssfont{m@thit}{\fontn@me}%
    \get@font{mthsy}{\math@syfnt}{\s@ze}\set@skchar{'60}%
    \ass@tfont{m@thsy}{\fontn@me}%
    \get@font{mthsy}{\math@syfnt}{\ss@ze}\set@skchar{'60}%
    \ass@sfont{m@thsy}{\fontn@me}%
    \get@font{mthsy}{\math@syfnt}{\sss@ze}\set@skchar{'60}%
    \ass@ssfont{m@thsy}{\fontn@me}%
    \math@atom{#1}{%
    \mathchoice%
      {\hbox{$\m@th\displaystyle#1$}}%
      {\hbox{$\m@th\textstyle#1$}}%
      {\hbox{$\m@th\scriptstyle#1$}}%
      {\hbox{$\m@th\scriptscriptstyle#1$}}}%
  \endgroup
  }%
}



\def\diameter{{\ifmmode\mathchoice
{\ooalign{\hfil\hbox{$\displaystyle/$}\hfil\crcr
{\hbox{$\displaystyle\mathchar"20D$}}}}
{\ooalign{\hfil\hbox{$\textstyle/$}\hfil\crcr
{\hbox{$\textstyle\mathchar"20D$}}}}
{\ooalign{\hfil\hbox{$\scriptstyle/$}\hfil\crcr
{\hbox{$\scriptstyle\mathchar"20D$}}}}
{\ooalign{\hfil\hbox{$\scriptscriptstyle/$}\hfil\crcr
{\hbox{$\scriptscriptstyle\mathchar"20D$}}}}
\else{\ooalign{\hfil/\hfil\crcr\mathhexbox20D}}%
\fi}}

\def\sq{\ifmmode\squareforqed\else{\unskip\nobreak\hfil
\penalty50\hskip1em\null\nobreak\hfil\squareforqed
\parfillskip=0pt\finalhyphendemerits=0\endgraf}\fi}
\def\squareforqed{\hbox{\rlap{$\sqcap$}$\sqcup$}}


\def\bbbc{{\mathchoice {\setbox0=\hbox{$\displaystyle\rm C$}\hbox{\hbox
to0pt{\kern0.4\wd0\vrule height0.9\ht0\hss}\box0}}
{\setbox0=\hbox{$\textstyle\rm C$}\hbox{\hbox
to0pt{\kern0.4\wd0\vrule height0.9\ht0\hss}\box0}}
{\setbox0=\hbox{$\scriptstyle\rm C$}\hbox{\hbox
to0pt{\kern0.4\wd0\vrule height0.9\ht0\hss}\box0}}
{\setbox0=\hbox{$\scriptscriptstyle\rm C$}\hbox{\hbox
to0pt{\kern0.4\wd0\vrule height0.9\ht0\hss}\box0}}}}
\def\bbbq{{\mathchoice {\setbox0=\hbox{$\displaystyle\rm
Q$}\hbox{\raise
0.15\ht0\hbox to0pt{\kern0.4\wd0\vrule height0.8\ht0\hss}\box0}}
{\setbox0=\hbox{$\textstyle\rm Q$}\hbox{\raise
0.15\ht0\hbox to0pt{\kern0.4\wd0\vrule height0.8\ht0\hss}\box0}}
{\setbox0=\hbox{$\scriptstyle\rm Q$}\hbox{\raise
0.15\ht0\hbox to0pt{\kern0.4\wd0\vrule height0.7\ht0\hss}\box0}}
{\setbox0=\hbox{$\scriptscriptstyle\rm Q$}\hbox{\raise
0.15\ht0\hbox to0pt{\kern0.4\wd0\vrule height0.7\ht0\hss}\box0}}}}
\def\bbbt{{\mathchoice {\setbox0=\hbox{$\displaystyle\rm
T$}\hbox{\hbox to0pt{\kern0.3\wd0\vrule height0.9\ht0\hss}\box0}}
{\setbox0=\hbox{$\textstyle\rm T$}\hbox{\hbox
to0pt{\kern0.3\wd0\vrule height0.9\ht0\hss}\box0}}
{\setbox0=\hbox{$\scriptstyle\rm T$}\hbox{\hbox
to0pt{\kern0.3\wd0\vrule height0.9\ht0\hss}\box0}}
{\setbox0=\hbox{$\scriptscriptstyle\rm T$}\hbox{\hbox
to0pt{\kern0.3\wd0\vrule height0.9\ht0\hss}\box0}}}}
\def\bbbs{{\mathchoice
{\setbox0=\hbox{$\displaystyle     \rm S$}\hbox{\raise0.5\ht0\hbox
to0pt{\kern0.35\wd0\vrule height0.45\ht0\hss}\hbox
to0pt{\kern0.55\wd0\vrule height0.5\ht0\hss}\box0}}
{\setbox0=\hbox{$\textstyle        \rm S$}\hbox{\raise0.5\ht0\hbox
to0pt{\kern0.35\wd0\vrule height0.45\ht0\hss}\hbox
to0pt{\kern0.55\wd0\vrule height0.5\ht0\hss}\box0}}
{\setbox0=\hbox{$\scriptstyle      \rm S$}\hbox{\raise0.5\ht0\hbox
to0pt{\kern0.35\wd0\vrule height0.45\ht0\hss}\raise0.05\ht0\hbox
to0pt{\kern0.5\wd0\vrule height0.45\ht0\hss}\box0}}
{\setbox0=\hbox{$\scriptscriptstyle\rm S$}\hbox{\raise0.5\ht0\hbox
to0pt{\kern0.4\wd0\vrule height0.45\ht0\hss}\raise0.05\ht0\hbox
to0pt{\kern0.55\wd0\vrule height0.45\ht0\hss}\box0}}}}
\def\bbbz{{\mathchoice {\hbox{$\sf\textstyle Z\kern-0.4em Z$}}
{\hbox{$\sf\textstyle Z\kern-0.4em Z$}}
{\hbox{$\sf\scriptstyle Z\kern-0.3em Z$}}
{\hbox{$\sf\scriptscriptstyle Z\kern-0.2em Z$}}}}


\def\Nulle{0} 
\def\Afe{1}   
\def\Hae{2}   
\def\Hbe{3}   
\def\Hce{4}   
\def\Hde{5}   


\newcount\LastMac       \LastMac=\Nulle

\newskip\half      \half=5.5pt plus 1.5pt minus 2.25pt
\newskip\one       \one=11pt plus 3pt minus 5.5pt
\newskip\onehalf   \onehalf=16.5pt plus 5.5pt minus 8.25pt
\newskip\two       \two=22pt plus 5.5pt minus 11pt

\def\Half{\addvspace{\half}}
\def\One{\addvspace{\one}}
\def\OneHalf{\addvspace{\onehalf}}
\def\Two{\addvspace{\two}}

\def\Raggedright{
  \rightskip=\z@ plus \hsize\relax
}

\def\Fullout{
  \rightskip=\z@\relax
}

\def\Hang#1#2{
  \hangindent=#1%
  \hangafter=#2\relax
}


\newif\ifsp@page
\def\pagestyle#1{\csname ps@#1\endcsname}
\def\thispagestyle#1{\global\sp@pagetrue\gdef\sp@type{#1}}

\def\ps@titlepage{%
  \def\@oddhead{\eightpoint\noindent \the\CatchLine
    \ifprod@font\else\qquad Printed\ \today\qquad
      (MN plain \TeX\ macros\ v\@version)\fi \hfil}%
  \let\@evenhead=\@oddhead
  \def\@oddfoot{\eightpoint\copyright\ \@pubyear\ RAS\hfil}%
  \def\@evenfoot{\hfil\eightpoint\noindent\copyright\ \@pubyear\ RAS}%
}

\def\ps@headings{%
  \def\@oddhead{\elevenpoint\it\noindent
    \hfill\the\RightHeader\hskip1.5em\rm\folio}%
  \def\@evenhead{\elevenpoint\noindent
    \folio\hskip1.5em\it\the\LeftHeader\hfill}%
  \def\@oddfoot{\eightpoint\noindent\copyright\ \@pubyear\ RAS,
    MNRAS {\bf \@volume}, \@pagerange\hfil}%
  \def\@evenfoot{\hfil\eightpoint\copyright\ \@pubyear\ RAS,
    MNRAS {\bf \@volume}, \@pagerange}%
}

\def\ps@plate{%
  \def\@oddhead{\eightpoint\noindent\plt@cap\hfil}%
  \def\@evenhead{\eightpoint\noindent\plt@cap\hfil}%
  \def\@oddfoot{\eightpoint\noindent\copyright\ \@pubyear\ RAS,
    MNRAS {\bf \@volume}, \@pagerange\hfil}%
  \def\@evenfoot{\hfil\eightpoint\copyright\ \@pubyear\ RAS,
    MNRAS {\bf \@volume}, \@pagerange}%
}



\def\title#1{
  \bgroup
    \vbox to 8pt{\vss}%
    \seventeenpoint
    \Raggedright
    \noindent \strut{\bf #1}\par
  \egroup
}

\def\author#1{
  \bgroup
    \ifnum\LastMac=\Afe \OneHalf\else \vskip 21pt\fi
    \fourteenpoint
    \Raggedright
    \noindent \strut #1\par
    \vskip 3pt%
  \egroup
}

\def\affiliation#1{
  \bgroup
    \vskip -4pt%
    \eightpoint
    \Raggedright
    \noindent \strut {\it #1}\par
  \egroup
  \LastMac=\Afe\relax
}

\def\acceptedline#1{
  \bgroup
    \Two
    \eightpoint
    \Raggedright
    \noindent \strut #1\par
  \egroup
}

\long\def\abstract#1{%
  \bgroup
    \vskip 20pt%
    \leftskip 11pc\rightskip\z@
    \noindent{\ninebf ABSTRACT}\par
    \tenpoint
    \Fullout
    \noindent #1\par
  \egroup
}

\long\def\keywords#1{
  \bgroup
    \Half
    \leftskip 11pc\rightskip\z@
    \tenpoint
    \Fullout
    \noindent\hbox{\bf Key words:}\ #1\par
  \egroup
}


\def\maketitle{%
  \EndOpening
  \ifsinglecol \else \MakePage\fi
}


\def\pageoffset#1#2{\hoffset=#1\relax\voffset=#2\relax}


\def\@nameuse#1{\csname #1\endcsname}
\def\arabic#1{\@arabic{\@nameuse{#1}}}
\def\alph#1{\@alph{\@nameuse{#1}}}
\def\Alph#1{\@Alph{\@nameuse{#1}}}
\def\@arabic#1{\number #1}
\def\@Alph#1{\ifcase#1\or A\or B\or C\or D\else\@Ialph{#1}\fi}
\def\@Ialph#1{\ifcase#1\or \or \or \or \or E\or F\or G\or H\or I\or J\or
   K\or L\or M\or N\or O\or P\or Q\or R\or S\or T\or U\or V\or W\or X\or
   Y\or Z\else\errmessage{Counter out of range}\fi}
\def\@alph#1{\ifcase#1\or a\or b\or c\or d\else\@ialph{#1}\fi}
\def\@ialph#1{\ifcase#1\or \or \or \or \or e\or f\or g\or h\or i\or j\or
   k\or l\or m\or n\or o\or p\or q\or r\or s\or t\or u\or v\or w\or x\or y\or
   z\else\errmessage{Counter out of range}\fi}


\newcount\Eqnno
\newcount\SubEqnno

\def\theeq{\arabic{Eqnno}}
\def\thesubeq{\alph{SubEqnno}}

\def\stepeq{\relax
  \global\SubEqnno \z@
  \global\advance\Eqnno \@ne\relax
  {\rm (\theeq)}%
}

\def\startsubeq{\relax
  \global\SubEqnno \z@
  \global\advance\Eqnno \@ne\relax
  \stepsubeq
}

\def\stepsubeq{\relax
  \global\advance\SubEqnno \@ne\relax
  {\rm (\theeq\thesubeq)}%
}


\newcount\Sec        
\newcount\SecSec
\newcount\SecSecSec

\def\thesection{\arabic{Sec}}
\def\thesubsection{\thesection.\arabic{SecSec}}
\def\thesubsubsection{\thesubsection.\arabic{SecSecSec}}

\Sec=\z@

\def\:{\let\@sptoken= } \:  
\def\:{\@xifnch} \expandafter\def\: {\futurelet\@tempc\@ifnch}

\def\@ifnextchar#1#2#3{%
  \let\@tempMACe #1%
  \def\@tempMACa{#2}%
  \def\@tempMACb{#3}%
  \futurelet \@tempMACc\@ifnch%
}

\def\@ifnch{%
\ifx \@tempMACc \@sptoken%
  \let\@tempMACd\@xifnch%
\else%
  \ifx \@tempMACc \@tempMACe%
    \let\@tempMACd\@tempMACa%
  \else%
    \let\@tempMACd\@tempMACb%
  \fi%
\fi%
\@tempMACd%
}

\def\@ifstar#1#2{\@ifnextchar *{\def\@tempMACa*{#1}\@tempMACa}{#2}}

\newskip\@tempskipb

\def\addvspace#1{%
  \ifvmode\else \endgraf\fi%
  \ifdim\lastskip=\z@%
    \vskip #1\relax%
  \else%
    \@tempskipb#1\relax\@xaddvskip%
  \fi%
}

\def\@xaddvskip{%
  \ifdim\lastskip<\@tempskipb%
    \vskip-\lastskip%
    \vskip\@tempskipb\relax%
  \else%
    \ifdim\@tempskipb<\z@%
      \ifdim\lastskip<\z@ \else%
        \advance\@tempskipb\lastskip%
        \vskip-\lastskip\vskip\@tempskipb%
      \fi%
    \fi%
  \fi%
}

\newskip\@tmpSKIP

\def\addpen#1{%
  \ifvmode
    \if@nobreak
    \else
      \ifdim\lastskip=\z@
        \penalty#1\relax
      \else
        \@tmpSKIP=\lastskip
        \vskip -\lastskip
        \penalty#1\vskip\@tmpSKIP
      \fi
    \fi
  \fi
}

\newcount\@clubpen   \@clubpen=\clubpenalty
\newif\if@nobreak    \@nobreakfalse

\def\@noafterindent{%
  \global\@nobreaktrue
  \everypar{\if@nobreak
              \global\@nobreakfalse
              \clubpenalty \@M
              {\setbox\z@\lastbox}%
              \LastMac=\Nulle\relax%
            \else
              \clubpenalty \@clubpen
              \everypar{}%
            \fi}%
}

\newcount\gds@cbrk   \gds@cbrk=-300

\def\@nohdbrk{\interlinepenalty \@M\relax}

\let\@par=\par
\def\@restorepar{\def\par{\@par}}

\newif\if@endpe   \@endpefalse
 
\def\@doendpe{\@endpetrue \@nobreakfalse \LastMac=\Nulle\relax%
     \def\par{\@restorepar\everypar{}\par\@endpefalse}%
              \everypar{\setbox\z@\lastbox\everypar{}\@endpefalse}%
}

\def\section{\@ifstar{\@ssection}{\@section}}

\def\@section#1{
  \if@nobreak
    \everypar{}%
    \ifnum\LastMac=\Hae \addvspace{\half}\fi
  \else
    \addpen{\gds@cbrk}%
    \addvspace{\two}%
  \fi
  \bgroup
    \ninepoint\bf
    \Raggedright
    \global\advance\Sec \@ne
    \ifappendix
      \global\Eqnno=\z@ \global\SubEqnno=\z@\relax
      \def\ch@ck{#1}%
      \ifx\ch@ck\empty \def\c@lon{}\else\def\c@lon{:}\fi
      \noindent\@nohdbrk APPENDIX\ \thesection\c@lon\hskip 0.5em%
        \uppercase{#1}\par
    \else
      \noindent\@nohdbrk\thesection\hskip 1pc \uppercase{#1}\par
    \fi
    \global\SecSec=\z@
  \egroup
  \nobreak
  \vskip\half
  \nobreak
  \@noafterindent
  \LastMac=\Hae\relax
}

\def\@ssection#1{
  \if@nobreak
    \everypar{}%
    \ifnum\LastMac=\Hae \addvspace{\half}\fi
  \else
    \addpen{\gds@cbrk}%
    \addvspace{\two}%
  \fi
  \bgroup
    \ninepoint\bf
    \Raggedright
    \noindent\@nohdbrk\uppercase{#1}\par
  \egroup
  \nobreak
  \vskip\half
  \nobreak
  \@noafterindent
  \LastMac=\Hae\relax
}

\def\subsection{\@ifstar{\@ssubsection}{\@subsection}}

\def\@subsection#1{
  \if@nobreak
    \everypar{}%
    \ifnum\LastMac=\Hae \addvspace{1pt plus 1pt minus .5pt}\fi
  \else
    \addpen{\gds@cbrk}%
    \addvspace{\onehalf}%
  \fi
  \bgroup
    \ninepoint\bf
    \Raggedright
    \global\advance\SecSec \@ne
    \noindent\@nohdbrk\thesubsection \hskip 1pc\relax #1\par
    \global\SecSecSec=\z@
  \egroup
  \nobreak
  \vskip\half
  \nobreak
  \@noafterindent
  \LastMac=\Hbe\relax
}

\def\@ssubsection#1{
  \if@nobreak
    \everypar{}%
    \ifnum\LastMac=\Hae \addvspace{1pt plus 1pt minus .5pt}\fi
  \else
    \addpen{\gds@cbrk}%
    \addvspace{\onehalf}%
  \fi
  \bgroup
    \ninepoint\bf
    \Raggedright
    \noindent\@nohdbrk #1\par
  \egroup
  \nobreak
  \vskip\half
  \nobreak
  \@noafterindent
  \LastMac=\Hbe\relax
}

\def\subsubsection{\@ifstar{\@ssubsubsection}{\@subsubsection}}

\def\@subsubsection#1{
  \if@nobreak
    \everypar{}%
    \ifnum\LastMac=\Hbe \addvspace{1pt plus 1pt minus .5pt}\fi
  \else
    \addpen{\gds@cbrk}%
    \addvspace{\onehalf}%
  \fi
  \bgroup
    \ninepoint\it
    \Raggedright
    \global\advance\SecSecSec \@ne
    \noindent\@nohdbrk\thesubsubsection \hskip 1pc\relax #1\par
  \egroup
  \nobreak
  \vskip\half
  \nobreak
  \@noafterindent
  \LastMac=\Hce\relax
}

\def\@ssubsubsection#1{
  \if@nobreak
    \everypar{}%
    \ifnum\LastMac=\Hbe \addvspace{1pt plus 1pt minus .5pt}\fi
  \else
    \addpen{\gds@cbrk}%
    \addvspace{\onehalf}%
  \fi
  \bgroup
    \ninepoint\it
    \Raggedright
    \noindent\@nohdbrk #1\par
  \egroup
  \nobreak
  \vskip\half
  \nobreak
  \@noafterindent
  \LastMac=\Hce\relax
}

\def\paragraph#1{
  \if@nobreak
    \everypar{}%
  \else
    \addpen{\gds@cbrk}%
    \addvspace{\one}%
  \fi%
  \bgroup%
    \ninepoint\it
    \noindent #1\ \nobreak%
  \egroup
  \LastMac=\Hde\relax
  \ignorespaces
}


\newif\ifappendix

\def\appendix{%
  \global\appendixtrue
  \def\thesection{\Alph{Sec}}%
  \def\thesubsection{\thesection\arabic{SecSec}}%
  \def\theeq{\thesection\arabic{Eqnno}}%
  \Sec=\z@ \SecSec=\z@ \SecSecSec=\z@ \Eqnno=\z@ \SubEqnno=\z@\relax
}




\def\beginlist{%
  \par\if@nobreak \else\addvspace{\half}\fi%
  \bgroup%
    \ninepoint
    \let\item=\list@item%
}

\def\list@item{%
  \par\noindent\hskip 1em\relax%
  \ignorespaces%
}

\def\endlist{\par\egroup\addvspace{\half}\@doendpe}


\def\beginrefs{%
  \par
  \bgroup
    \eightpoint
    \Fullout
    \let\bibitem=\bib@item
}

\def\bib@item{%
  \par\parindent=1.5em\Hang{1.5em}{1}%
  \everypar={\Hang{1.5em}{1}\ignorespaces}%
  \noindent\ignorespaces
}

\def\endrefs{\par\egroup\@doendpe}


\newtoks\CatchLine

\def\@journal{Mon.\ Not.\ R.\ Astron.\ Soc.\ }  
\def\@pubyear{1994}        
\def\@pagerange{000--000}  
\def\@volume{000}          
\def\@microfiche{}         %

\def\pubyear#1{\gdef\@pubyear{#1}\@makecatchline}
\def\pagerange#1{\gdef\@pagerange{#1}\@makecatchline}
\def\volume#1{\gdef\@volume{#1}\@makecatchline}
\def\microfiche#1{\gdef\@microfiche{and Microfiche\ #1}\@makecatchline}

\def\@makecatchline{%
  \global\CatchLine{%
    {\rm \@journal {\bf \@volume},\ \@pagerange\ (\@pubyear)\ \@microfiche}}%
}

\@makecatchline 

\newtoks\LeftHeader
\def\shortauthor#1{
  \global\LeftHeader{#1}%
}

\newtoks\RightHeader
\def\shorttitle#1{
  \global\RightHeader{#1}%
}

\def\PageHead{
  \begingroup
    \ifsp@page
      \csname ps@\sp@type\endcsname
    \fi
    \ifodd\pageno
      \let\the@head=\@oddhead
    \else
      \let\the@head=\@evenhead
    \fi
    \vbox to \z@{\vskip-22.5\p@%
      \hbox to \PageWidth{\vbox to8.5\p@{}%
        \the@head
      }%
    \vss}%
  \endgroup
  \nointerlineskip
}

\gdef\PageFoot{%
  \nointerlineskip%
  \begingroup
  \ifsp@page
    \csname ps@\sp@type\endcsname
    \global\sp@pagefalse
  \fi
  \vbox to 22pt{\vfil%
    \hbox to \PageWidth{%
      \eightpoint\strut\noindent
      \ifodd\pageno
        \@oddfoot
      \else
        \@evenfoot
      \fi
    }%
  }%
  \endgroup
}

\def\today{%
  \number\day\space
  \ifcase\month\or January\or February\or March\or April\or May\or June\or
    July\or August\or September\or October\or November\or December\fi
  \space\number\year%
}

\def\authorcomment#1{%
  \gdef\PageFoot{%
    \nointerlineskip%
    \vbox to 20pt{\vfil%
      \hbox to \PageWidth{\elevenpoint\noindent \hfil #1 \hfil}}%
  }%
}


\newif\ifplate@page
\newbox\plt@box

\def\beginplatepage{%
  \let\plate=\plate@head
  \let\caption=\fig@caption
  \global\setbox\plt@box=\vbox\bgroup
  \TEMPDIMEN=\PageWidth 
  \hsize=\PageWidth\relax
}

\def\endplatepage{\par\egroup\global\plate@pagetrue}
\def\plate@head#1{\gdef\plt@cap{#1}}


\def\letters{%
  \gdef\folio{\ifnum\pageno<\z@ L\romannumeral-\pageno
    \else L\number\pageno \fi}%
}


\newdimen\mathindent

\global\mathindent=\z@
\global\everydisplay{\global\@dspwd=\displaywidth\displaysetup}


\def\@displaylines#1{
  {}$\displ@y\hbox{\vbox{\halign{$\@lign\hfil\displaystyle##\hfil$\crcr
  #1\crcr}}}${}%
}

\def\@eqalign#1{\null\vcenter{\openup\jot\m@th
  \ialign{\strut\hfil$\displaystyle{##}$&$\displaystyle{{}##}$\hfil
      \crcr#1\crcr}}%
}

\def\@eqalignno#1{
  \global\advance\@dspwd by -\mathindent%
  {}$\displ@y\hbox{\vbox{\halign to\@dspwd%
  {\hfil$\@lign\displaystyle{##}$\tabskip\z@skip
  &$\@lign\displaystyle{{}##}$\hfil\tabskip\centering
  &\llap{$\@lign##$}\tabskip\z@skip\crcr
  #1\crcr}}}${}%
}


\global\let\displaylines=\@displaylines
\global\let\eqalign=\@eqalign
\global\let\eqalignno=\@eqalignno
\global\let\leqalignno=\@eqalignno

\newdimen\@dspwd   \@dspwd=\z@
\newif\if@eqno
\newif\if@leqno
\newtoks\@eqn
\newtoks\@eq

\def\displaysetup#1$${\displaytest#1\eqno\eqno\displaytest}

\def\displaytest#1\eqno#2\eqno#3\displaytest{%
 \if!#3!\ldisplaytest#1\leqno\leqno\ldisplaytest
 \else\@eqnotrue\@leqnofalse\@eqn={#2}\@eq={#1}\fi
 \generaldisplay$$}

\def\ldisplaytest#1\leqno#2\leqno#3\ldisplaytest{%
\@eq={#1}%
 \if!#3!\@eqnofalse\else\@eqnotrue\@leqnotrue
  \@eqn={#2}\fi}

\def\generaldisplay{%
  \if@eqno
    \if@leqno
      \hbox to \displaywidth{\noindent
        \rlap{$\displaystyle\the\@eqn$}%
        \hskip\mathindent$\displaystyle\the\@eq$\hfil}%
    \else
      \hbox to \displaywidth{\noindent
        \hskip\mathindent
        $\displaystyle\the\@eq$\hfil$\displaystyle\the\@eqn$}%
    \fi
  \else
    \hbox to \displaywidth{\noindent
      \hskip\mathindent$\displaystyle\the\@eq$\hfil}%
  \fi
}


\def\@notice{%
  \par\Two%
  \noindent{\b@ls{11pt}\ninerm This paper has been produced using the
    Royal Astronomical Society/Blackwell Science \TeX\ macros.\par}%
}

\outer\def\bye{\@notice\par\vfill\supereject\end}


\def\start@mess{%
  Monthly notices of the RAS journal style (\@typeface)\space
    v\@version,\space \@verdate.%
}

\everyjob{\Warn{\start@mess}}



\newif\if@debug \@debugfalse  

\def\Print#1{\if@debug\immediate\write16{#1}\else \fi}
\def\Warn#1{\immediate\write16{#1}}
\def\wlog#1{}

\newcount\Iteration 

\def\Single{0} \def\Double{1}                 
\def\Figure{0} \def\Table{1}                  

\def\InStack{0}  
\def\InZoneA{1}
\def\InZoneB{2}
\def\InZoneC{3}

\newcount\TEMPCOUNT 
\newdimen\TEMPDIMEN 
\newbox\TEMPBOX     
\newbox\VOIDBOX     

\newcount\LengthOfStack 
\newcount\MaxItems      
\newcount\StackPointer
\newcount\Point         
\newcount\NextFigure    
\newcount\NextTable     
\newcount\NextItem      

\newcount\StatusStack   
\newcount\NumStack      
\newcount\TypeStack     
\newcount\SpanStack     
\newcount\BoxStack      

\newcount\ItemSTATUS    
\newcount\ItemNUMBER    
\newcount\ItemTYPE      
\newcount\ItemSPAN      
\newbox\ItemBOX         
\newdimen\ItemSIZE      

\newdimen\PageHeight    
\newdimen\TextLeading   
\newdimen\Feathering    
\newcount\LinesPerPage  
\newdimen\ColumnWidth   
\newdimen\ColumnGap     
\newdimen\PageWidth     
\newdimen\BodgeHeight   
\newcount\Leading       

\newdimen\ZoneBSize  
\newdimen\TextSize   
\newbox\ZoneABOX     
\newbox\ZoneBBOX     
\newbox\ZoneCBOX     

\newif\ifFirstSingleItem
\newif\ifFirstZoneA
\newif\ifMakePageInComplete
\newif\ifMoreFigures \MoreFiguresfalse 
\newif\ifMoreTables  \MoreTablesfalse  

\newif\ifFigInZoneB 
\newif\ifFigInZoneC 
\newif\ifTabInZoneB 
\newif\ifTabInZoneC

\newif\ifZoneAFullPage

\newbox\MidBOX    
\newbox\LeftBOX
\newbox\RightBOX
\newbox\PageBOX   

\newif\ifLeftCOL  
\LeftCOLtrue

\newdimen\ZoneBAdjust

\newcount\ItemFits
\def\Yes{1}
\def\No{2}


\MaxItems=15
\NextFigure=\z@        
\NextTable=\@ne

\BodgeHeight=6pt
\TextLeading=11pt    
\Leading=11
\Feathering=\z@      
\LinesPerPage=61     
\topskip=\TextLeading
\ColumnWidth=20pc    
\ColumnGap=2pc       

\newskip\ItemSepamount  
\ItemSepamount=\TextLeading plus \TextLeading minus 4pt

\parskip=\z@ plus .1pt
\parindent=18pt
\widowpenalty=\z@
\clubpenalty=10000
\tolerance=1500
\hbadness=1500
\abovedisplayskip=6pt plus 2pt minus 1pt
\belowdisplayskip=6pt plus 2pt minus 1pt
\abovedisplayshortskip=6pt plus 2pt minus 1pt
\belowdisplayshortskip=6pt plus 2pt minus 1pt

\frenchspacing

\ninepoint 

\PageHeight=682pt
\PageWidth=2\ColumnWidth
\advance\PageWidth by \ColumnGap

\pagestyle{headings}




\newcount\DUMMY \StatusStack=\allocationnumber
\newcount\DUMMY \newcount\DUMMY \newcount\DUMMY 
\newcount\DUMMY \newcount\DUMMY \newcount\DUMMY 
\newcount\DUMMY \newcount\DUMMY \newcount\DUMMY
\newcount\DUMMY \newcount\DUMMY \newcount\DUMMY 
\newcount\DUMMY \newcount\DUMMY \newcount\DUMMY

\newcount\DUMMY \NumStack=\allocationnumber
\newcount\DUMMY \newcount\DUMMY \newcount\DUMMY 
\newcount\DUMMY \newcount\DUMMY \newcount\DUMMY 
\newcount\DUMMY \newcount\DUMMY \newcount\DUMMY 
\newcount\DUMMY \newcount\DUMMY \newcount\DUMMY 
\newcount\DUMMY \newcount\DUMMY \newcount\DUMMY

\newcount\DUMMY \TypeStack=\allocationnumber
\newcount\DUMMY \newcount\DUMMY \newcount\DUMMY 
\newcount\DUMMY \newcount\DUMMY \newcount\DUMMY 
\newcount\DUMMY \newcount\DUMMY \newcount\DUMMY 
\newcount\DUMMY \newcount\DUMMY \newcount\DUMMY 
\newcount\DUMMY \newcount\DUMMY \newcount\DUMMY

\newcount\DUMMY \SpanStack=\allocationnumber
\newcount\DUMMY \newcount\DUMMY \newcount\DUMMY 
\newcount\DUMMY \newcount\DUMMY \newcount\DUMMY 
\newcount\DUMMY \newcount\DUMMY \newcount\DUMMY 
\newcount\DUMMY \newcount\DUMMY \newcount\DUMMY 
\newcount\DUMMY \newcount\DUMMY \newcount\DUMMY

\newbox\DUMMY   \BoxStack=\allocationnumber
\newbox\DUMMY   \newbox\DUMMY \newbox\DUMMY 
\newbox\DUMMY   \newbox\DUMMY \newbox\DUMMY 
\newbox\DUMMY   \newbox\DUMMY \newbox\DUMMY 
\newbox\DUMMY   \newbox\DUMMY \newbox\DUMMY 
\newbox\DUMMY   \newbox\DUMMY \newbox\DUMMY

\def\wlog{\immediate\write\m@ne}


\def\GetItemAll#1{%
 \GetItemSTATUS{#1}
 \GetItemNUMBER{#1}
 \GetItemTYPE{#1}
 \GetItemSPAN{#1}
 \GetItemBOX{#1}
}

\def\GetItemSTATUS#1{%
 \Point=\StatusStack
 \advance\Point by #1
 \global\ItemSTATUS=\count\Point
}

\def\GetItemNUMBER#1{%
 \Point=\NumStack
 \advance\Point by #1
 \global\ItemNUMBER=\count\Point
}

\def\GetItemTYPE#1{%
 \Point=\TypeStack
 \advance\Point by #1
 \global\ItemTYPE=\count\Point
}

\def\GetItemSPAN#1{%
 \Point\SpanStack
 \advance\Point by #1
 \global\ItemSPAN=\count\Point
}

\def\GetItemBOX#1{%
 \Point=\BoxStack
 \advance\Point by #1
 \global\setbox\ItemBOX=\vbox{\copy\Point}
 \global\ItemSIZE=\ht\ItemBOX
 \global\advance\ItemSIZE by \dp\ItemBOX
 \TEMPCOUNT=\ItemSIZE
 \divide\TEMPCOUNT by \Leading
 \divide\TEMPCOUNT by 65536
 \advance\TEMPCOUNT \@ne
 \ItemSIZE=\TEMPCOUNT pt
 \global\multiply\ItemSIZE by \Leading
}


\def\JoinStack{%
 \ifnum\LengthOfStack=\MaxItems 
  \Warn{WARNING: Stack is full...some items will be lost!}
 \else
  \Point=\StatusStack
  \advance\Point by \LengthOfStack
  \global\count\Point=\ItemSTATUS
  \Point=\NumStack
  \advance\Point by \LengthOfStack
  \global\count\Point=\ItemNUMBER
  \Point=\TypeStack
  \advance\Point by \LengthOfStack
  \global\count\Point=\ItemTYPE
  \Point\SpanStack
  \advance\Point by \LengthOfStack
  \global\count\Point=\ItemSPAN
  \Point=\BoxStack
  \advance\Point by \LengthOfStack
  \global\setbox\Point=\vbox{\copy\ItemBOX}
  \global\advance\LengthOfStack \@ne
  \ifnum\ItemTYPE=\Figure 
   \global\MoreFigurestrue
  \else
   \global\MoreTablestrue
  \fi
 \fi
}


\def\LeaveStack#1{%
 {\Iteration=#1
 \loop
 \ifnum\Iteration<\LengthOfStack
  \advance\Iteration \@ne
  \GetItemSTATUS{\Iteration}
   \advance\Point by \m@ne
   \global\count\Point=\ItemSTATUS
  \GetItemNUMBER{\Iteration}
   \advance\Point by \m@ne
   \global\count\Point=\ItemNUMBER
  \GetItemTYPE{\Iteration}
   \advance\Point by \m@ne
   \global\count\Point=\ItemTYPE
  \GetItemSPAN{\Iteration}
   \advance\Point by \m@ne
   \global\count\Point=\ItemSPAN
  \GetItemBOX{\Iteration}
   \advance\Point by \m@ne
   \global\setbox\Point=\vbox{\copy\ItemBOX}
 \repeat}
 \global\advance\LengthOfStack by \m@ne
}


\newif\ifStackNotClean

\def\CleanStack{%
 \StackNotCleantrue
 {\Iteration=\z@
  \loop
   \ifStackNotClean
    \GetItemSTATUS{\Iteration}
    \ifnum\ItemSTATUS=\InStack
     \advance\Iteration \@ne
     \else
      \LeaveStack{\Iteration}
    \fi
   \ifnum\LengthOfStack<\Iteration
    \StackNotCleanfalse
   \fi
 \repeat}
}


\def\FindItem#1#2{%
 \global\StackPointer=\m@ne 
 {\Iteration=\z@
  \loop
  \ifnum\Iteration<\LengthOfStack
   \GetItemSTATUS{\Iteration}
   \ifnum\ItemSTATUS=\InStack
    \GetItemTYPE{\Iteration}
    \ifnum\ItemTYPE=#1
     \GetItemNUMBER{\Iteration}
     \ifnum\ItemNUMBER=#2
      \global\StackPointer=\Iteration
      \Iteration=\LengthOfStack 
     \fi
    \fi
   \fi
  \advance\Iteration \@ne
 \repeat}
}


\def\FindNext{%
 \global\StackPointer=\m@ne 
 {\Iteration=\z@
  \loop
  \ifnum\Iteration<\LengthOfStack
   \GetItemSTATUS{\Iteration}
   \ifnum\ItemSTATUS=\InStack
    \GetItemTYPE{\Iteration}
   \ifnum\ItemTYPE=\Figure
    \ifMoreFigures
      \global\NextItem=\Figure
      \global\StackPointer=\Iteration
      \Iteration=\LengthOfStack 
    \fi
   \fi
   \ifnum\ItemTYPE=\Table
    \ifMoreTables
      \global\NextItem=\Table
      \global\StackPointer=\Iteration
      \Iteration=\LengthOfStack 
    \fi
   \fi
  \fi
  \advance\Iteration \@ne
 \repeat}
}


\def\ChangeStatus#1#2{%
 \Point=\StatusStack
 \advance\Point by #1
 \global\count\Point=#2
}



\def\Zone{\InZoneA}

\ZoneBAdjust=\z@

\def\MakePage{
 \global\ZoneBSize=\PageHeight
 \global\TextSize=\ZoneBSize
 \global\ZoneAFullPagefalse
 \global\topskip=\TextLeading
 \MakePageInCompletetrue
 \MoreFigurestrue
 \MoreTablestrue
 \FigInZoneBfalse
 \FigInZoneCfalse
 \TabInZoneBfalse
 \TabInZoneCfalse
 \global\FirstSingleItemtrue
 \global\FirstZoneAtrue
 \global\setbox\ZoneABOX=\box\VOIDBOX
 \global\setbox\ZoneBBOX=\box\VOIDBOX
 \global\setbox\ZoneCBOX=\box\VOIDBOX
 \loop
  \ifMakePageInComplete
 \FindNext
 \ifnum\StackPointer=\m@ne
  \NextItem=\m@ne
  \MoreFiguresfalse
  \MoreTablesfalse
 \fi
 \ifnum\NextItem=\Figure
   \FindItem{\Figure}{\NextFigure}
   \ifnum\StackPointer=\m@ne \global\MoreFiguresfalse
   \else
    \GetItemSPAN{\StackPointer}
    \ifnum\ItemSPAN=\Single \def\Zone{\InZoneB}\relax
     \ifFigInZoneC \global\MoreFiguresfalse\fi
    \else
     \def\Zone{\InZoneA}
     \ifFigInZoneB \def\Zone{\InZoneC}\fi
    \fi
   \fi
   \ifMoreFigures\Print{}\FigureItems\fi
 \fi
\ifnum\NextItem=\Table
   \FindItem{\Table}{\NextTable}
   \ifnum\StackPointer=\m@ne \global\MoreTablesfalse
   \else
    \GetItemSPAN{\StackPointer}
    \ifnum\ItemSPAN=\Single\relax
     \ifTabInZoneC \global\MoreTablesfalse\fi
    \else
     \def\Zone{\InZoneA}
     \ifTabInZoneB \def\Zone{\InZoneC}\fi
    \fi
   \fi
   \ifMoreTables\Print{}\TableItems\fi
 \fi
   \MakePageInCompletefalse 
   \ifMoreFigures\MakePageInCompletetrue\fi
   \ifMoreTables\MakePageInCompletetrue\fi
 \repeat
 \ifZoneAFullPage
  \global\TextSize=\z@
  \global\ZoneBSize=\z@
  \global\vsize=\z@\relax
  \global\topskip=\z@\relax
  \vbox to \z@{\vss}
  \eject
 \else
 \global\advance\ZoneBSize by -\ZoneBAdjust
 \global\vsize=\ZoneBSize
 \global\hsize=\ColumnWidth
 \global\ZoneBAdjust=\z@
 \ifdim\TextSize<23pt
 \Warn{}
 \Warn{* Making column fall short: TextSize=\the\TextSize *}
 \vskip-\lastskip\eject\fi
 \fi
}

\def\MakeRightCol{
 \global\TextSize=\ZoneBSize
 \MakePageInCompletetrue
 \MoreFigurestrue
 \MoreTablestrue
 \global\FirstSingleItemtrue
 \global\setbox\ZoneBBOX=\box\VOIDBOX
 \def\Zone{\InZoneB}
 \loop
  \ifMakePageInComplete
 \FindNext
 \ifnum\StackPointer=\m@ne
  \NextItem=\m@ne
  \MoreFiguresfalse
  \MoreTablesfalse
 \fi
 \ifnum\NextItem=\Figure
   \FindItem{\Figure}{\NextFigure}
   \ifnum\StackPointer=\m@ne \MoreFiguresfalse
   \else
    \GetItemSPAN{\StackPointer}
    \ifnum\ItemSPAN=\Double\relax
     \MoreFiguresfalse\fi
   \fi
   \ifMoreFigures\Print{}\FigureItems\fi
 \fi
 \ifnum\NextItem=\Table
   \FindItem{\Table}{\NextTable}
   \ifnum\StackPointer=\m@ne \MoreTablesfalse
   \else
    \GetItemSPAN{\StackPointer}
    \ifnum\ItemSPAN=\Double\relax
     \MoreTablesfalse\fi
   \fi
   \ifMoreTables\Print{}\TableItems\fi
 \fi
   \MakePageInCompletefalse 
   \ifMoreFigures\MakePageInCompletetrue\fi
   \ifMoreTables\MakePageInCompletetrue\fi
 \repeat
 \ifZoneAFullPage
  \global\TextSize=\z@
  \global\ZoneBSize=\z@
  \global\vsize=\z@\relax
  \global\topskip=\z@\relax
  \vbox to \z@{\vss}
  \eject
 \else
 \global\vsize=\ZoneBSize
 \global\hsize=\ColumnWidth
 \ifdim\TextSize<23pt
 \Warn{}
 \Warn{* Making column fall short: TextSize=\the\TextSize *}
 \vskip-\lastskip\eject\fi
\fi
}

\def\FigureItems{
 \Print{Considering...}
 \ShowItem{\StackPointer}
 \GetItemBOX{\StackPointer} 
 \GetItemSPAN{\StackPointer}
  \CheckFitInZone 
  \ifnum\ItemFits=\Yes
   \ifnum\ItemSPAN=\Single
     \ChangeStatus{\StackPointer}{\InZoneB} 
     \global\FigInZoneBtrue
     \ifFirstSingleItem
      \hbox{}\vskip-\BodgeHeight
     \global\advance\ItemSIZE by \TextLeading
     \fi
     \unvbox\ItemBOX\ItemSep
     \global\FirstSingleItemfalse
     \global\advance\TextSize by -\ItemSIZE
     \global\advance\TextSize by -\TextLeading
   \else
    \ifFirstZoneA
     \global\advance\ItemSIZE by \TextLeading
     \global\FirstZoneAfalse\fi
    \global\advance\TextSize by -\ItemSIZE
    \global\advance\TextSize by -\TextLeading
    \global\advance\ZoneBSize by -\ItemSIZE
    \global\advance\ZoneBSize by -\TextLeading
    \ifFigInZoneB\relax
     \else
     \ifdim\TextSize<3\TextLeading
     \global\ZoneAFullPagetrue
     \fi
    \fi
    \ChangeStatus{\StackPointer}{\Zone}
    \ifnum\Zone=\InZoneC \global\FigInZoneCtrue\fi
  \fi
   \Print{TextSize=\the\TextSize}
   \Print{ZoneBSize=\the\ZoneBSize}
  \global\advance\NextFigure \@ne
   \Print{This figure has been placed.}
  \else
   \Print{No space available for this figure...holding over.}
   \Print{}
   \global\MoreFiguresfalse
  \fi
}

\def\TableItems{
 \Print{Considering...}
 \ShowItem{\StackPointer}
 \GetItemBOX{\StackPointer} 
 \GetItemSPAN{\StackPointer}
  \CheckFitInZone 
  \ifnum\ItemFits=\Yes
   \ifnum\ItemSPAN=\Single
    \ChangeStatus{\StackPointer}{\InZoneB}
     \global\TabInZoneBtrue
     \ifFirstSingleItem
      \hbox{}\vskip-\BodgeHeight
     \global\advance\ItemSIZE by \TextLeading
     \fi
     \unvbox\ItemBOX\ItemSep
     \global\FirstSingleItemfalse
     \global\advance\TextSize by -\ItemSIZE
     \global\advance\TextSize by -\TextLeading
   \else
    \ifFirstZoneA
    \global\advance\ItemSIZE by \TextLeading
    \global\FirstZoneAfalse\fi
    \global\advance\TextSize by -\ItemSIZE
    \global\advance\TextSize by -\TextLeading
    \global\advance\ZoneBSize by -\ItemSIZE
    \global\advance\ZoneBSize by -\TextLeading
    \ifFigInZoneB\relax
     \else
     \ifdim\TextSize<3\TextLeading
     \global\ZoneAFullPagetrue
     \fi
    \fi
    \ChangeStatus{\StackPointer}{\Zone}
    \ifnum\Zone=\InZoneC \global\TabInZoneCtrue\fi
   \fi
  \global\advance\NextTable \@ne
   \Print{This table has been placed.}
  \else
  \Print{No space available for this table...holding over.}
   \Print{}
   \global\MoreTablesfalse
  \fi
}


\def\CheckFitInZone{%
{\advance\TextSize by -\ItemSIZE
 \advance\TextSize by -\TextLeading
 \ifFirstSingleItem
  \advance\TextSize by \TextLeading
 \fi
 \ifnum\Zone=\InZoneA\relax
  \else \advance\TextSize by -\ZoneBAdjust
 \fi
 \ifdim\TextSize<3\TextLeading \global\ItemFits=\No
 \else \global\ItemFits=\Yes\fi}
}

\def\BeginOpening{%
  \ninepoint
  \thispagestyle{titlepage}%
  \global\setbox\ItemBOX=\vbox\bgroup%
    \hsize=\PageWidth%
    \hrule height \z@
    \ifsinglecol\vskip 6pt\fi 
}

\let\begintopmatter=\BeginOpening  

\def\EndOpening{%
  \One
  \egroup
  \ifsinglecol
    \box\ItemBOX%
    \vskip\TextLeading plus 2\TextLeading
    \@noafterindent
  \else
    \ItemNUMBER=\z@%
    \ItemTYPE=\Figure
    \ItemSPAN=\Double
    \ItemSTATUS=\InStack
    \JoinStack
  \fi
}


\newif\if@here  \@herefalse

\def\no@float{\global\@heretrue}
\let\nofloat=\relax 

\def\beginfigure{%
  \@ifstar{\global\@dfloattrue \@bfigure}{\global\@dfloatfalse \@bfigure}%
}

\def\@bfigure#1{%
  \par
  \if@dfloat
    \ItemSPAN=\Double
    \TEMPDIMEN=\PageWidth
  \else
    \ItemSPAN=\Single
    \TEMPDIMEN=\ColumnWidth
  \fi
  \ifsinglecol
    \TEMPDIMEN=\PageWidth
  \else
    \ItemSTATUS=\InStack
    \ItemNUMBER=#1%
    \ItemTYPE=\Figure
  \fi
  \bgroup
    \hsize=\TEMPDIMEN
    \global\setbox\ItemBOX=\vbox\bgroup
      \eightpoint\nostb@ls{10pt}%
      \let\caption=\fig@caption
      \ifsinglecol \let\nofloat=\no@float\fi
}

\def\fig@caption#1{%
  \vskip 5.5pt plus 6pt%
  \bgroup 
    \eightpoint\nostb@ls{10pt}%
    \setbox\TEMPBOX=\hbox{#1}%
    \ifdim\wd\TEMPBOX>\TEMPDIMEN
      \noindent \unhbox\TEMPBOX\par
    \else
      \hbox to \hsize{\hfil\unhbox\TEMPBOX\hfil}%
    \fi
  \egroup
}

\def\endfigure{%
  \par\egroup 
  \egroup
  \ifsinglecol
    \if@here \midinsert\global\@herefalse\else \topinsert\fi
      \unvbox\ItemBOX
    \endinsert
  \else
    \JoinStack
    \Print{Processing source for figure \the\ItemNUMBER}%
  \fi
}


\newbox\tab@cap@box
\def\tab@caption#1{\global\setbox\tab@cap@box=\hbox{#1\par}}

\newtoks\tab@txt@toks
\long\def\tab@txt#1{\global\tab@txt@toks={#1}\global\table@txttrue}

\newif\iftable@txt  \table@txtfalse
\newif\if@dfloat    \@dfloatfalse

\def\begintable{%
  \@ifstar{\global\@dfloattrue \@btable}{\global\@dfloatfalse \@btable}%
}

\def\@btable#1{%
  \par
  \if@dfloat
    \ItemSPAN=\Double
    \TEMPDIMEN=\PageWidth
  \else
    \ItemSPAN=\Single
    \TEMPDIMEN=\ColumnWidth
  \fi
  \ifsinglecol
    \TEMPDIMEN=\PageWidth
  \else
    \ItemSTATUS=\InStack
    \ItemNUMBER=#1%
    \ItemTYPE=\Table
  \fi
  \bgroup
    \eightpoint\nostb@ls{10pt}%
    \global\setbox\ItemBOX=\vbox\bgroup
      \let\caption=\tab@caption
      \let\tabletext=\tab@txt
      \ifsinglecol \let\nofloat=\no@float\fi
}

\def\endtable{%
  \par\egroup 
  \egroup
  \setbox\TEMPBOX=\hbox to \TEMPDIMEN{%
    \eightpoint\nostb@ls{10pt}%
    \hss
    \vbox{%
      \hsize=\wd\ItemBOX
      \ifvoid\tab@cap@box
      \else
        \noindent\unhbox\tab@cap@box
        \vskip 5.5pt plus 6pt%
      \fi
      \box\ItemBOX
      \iftable@txt
        \vskip 10pt%
        \noindent\the\tab@txt@toks
        \global\table@txtfalse
      \fi
    }%
    \hss
  }%
  \ifsinglecol
    \if@here \midinsert\global\@herefalse\else \topinsert\fi
      \box\TEMPBOX
    \endinsert
  \else
    \global\setbox\ItemBOX=\box\TEMPBOX
    \JoinStack
    \Print{Processing source for table \the\ItemNUMBER}%
  \fi
}

\def\UnloadZoneA{%
\FirstZoneAtrue
 \Iteration=\z@
  \loop
   \ifnum\Iteration<\LengthOfStack
    \GetItemSTATUS{\Iteration}
    \ifnum\ItemSTATUS=\InZoneA
     \GetItemBOX{\Iteration}
     \ifFirstZoneA \vbox to \BodgeHeight{\vfil}%
     \FirstZoneAfalse\fi
     \unvbox\ItemBOX\ItemSep
     \LeaveStack{\Iteration}
     \else
     \advance\Iteration \@ne
   \fi
 \repeat
}

\def\UnloadZoneC{%
\Iteration=\z@
  \loop
   \ifnum\Iteration<\LengthOfStack
    \GetItemSTATUS{\Iteration}
    \ifnum\ItemSTATUS=\InZoneC
     \GetItemBOX{\Iteration}
     \ItemSep\unvbox\ItemBOX
     \LeaveStack{\Iteration}
     \else
     \advance\Iteration \@ne
   \fi
 \repeat
}


\def\ShowItem#1{
  {\GetItemAll{#1}
  \Print{\the#1:
  {TYPE=\ifnum\ItemTYPE=\Figure Figure\else Table\fi}
  {NUMBER=\the\ItemNUMBER}
  {SPAN=\ifnum\ItemSPAN=\Single Single\else Double\fi}
  {SIZE=\the\ItemSIZE}}}
}

\def\ShowStack{%
 \Print{}
 \Print{LengthOfStack = \the\LengthOfStack}
 \ifnum\LengthOfStack=\z@ \Print{Stack is empty}\fi
 \Iteration=\z@
 \loop
 \ifnum\Iteration<\LengthOfStack
  \ShowItem{\Iteration}
  \advance\Iteration \@ne
 \repeat
}

\def\B#1#2{%
\hbox{\vrule\kern-0.4pt\vbox to #2{%
\hrule width #1\vfill\hrule}\kern-0.4pt\vrule}
}


\newif\ifsinglecol   \singlecolfalse

\def\onecolumn{%
  \global\output={\singlecoloutput}%
  \global\hsize=\PageWidth
  \global\vsize=\PageHeight
  \global\ColumnWidth=\hsize
  \global\TextLeading=12pt
  \global\Leading=12
  \global\singlecoltrue
  \global\let\onecolumn=\relax
  \global\let\footnote=\sing@footnote
  \global\let\vfootnote=\sing@vfootnote
  \ninepoint 
  \message{(Single column)}%
}

\def\singlecoloutput{%
  \shipout\vbox{\PageHead\vbox to \PageHeight{\pagebody\vss}\PageFoot}%
  \advancepageno
  \ifplate@page
    \shipout\vbox{%
      \sp@pagetrue
      \def\sp@type{plate}%
      \global\plate@pagefalse
      \PageHead\vbox to \PageHeight{\unvbox\plt@box\vfil}\PageFoot%
    }%
    \message{[plate]}%
    \advancepageno
  \fi
  \ifnum\outputpenalty>-\@MM \else\dosupereject\fi%
}

\def\ItemSep{\vskip\ItemSepamount\relax}

\def\ItemSepbreak{\par\ifdim\lastskip<\ItemSepamount
  \removelastskip\penalty-200\ItemSep\fi%
}


\let\@@endinsert=\endinsert 

\def\endinsert{\egroup 
  \if@mid \dimen@\ht\z@ \advance\dimen@\dp\z@ \advance\dimen@12\p@
    \advance\dimen@\pagetotal \advance\dimen@-\pageshrink
    \ifdim\dimen@>\pagegoal\@midfalse\p@gefalse\fi\fi
  \if@mid \ItemSep\box\z@\ItemSepbreak
  \else\insert\topins{\penalty100 
    \splittopskip\z@skip
    \splitmaxdepth\maxdimen \floatingpenalty\z@
    \ifp@ge \dimen@\dp\z@
    \vbox to\vsize{\unvbox\z@\kern-\dimen@}
    \else \box\z@\nobreak\ItemSep\fi}\fi\endgroup%
}


\def\gobbleone#1{}
\def\gobbletwo#1#2{}
\let\footnote=\gobbletwo 
\let\vfootnote=\gobbleone

\def\sing@footnote#1{\let\@sf\empty 
  \ifhmode\edef\@sf{\spacefactor\the\spacefactor}\/\fi
  \hbox{$^{\hbox{\eightpoint #1}}$}\@sf\sing@vfootnote{#1}%
}

\def\sing@vfootnote#1{\insert\footins\bgroup\eightpoint\b@ls{9pt}%
  \interlinepenalty\interfootnotelinepenalty
  \splittopskip\ht\strutbox 
  \splitmaxdepth\dp\strutbox \floatingpenalty\@MM
  \leftskip\z@skip \rightskip\z@skip \spaceskip\z@skip \xspaceskip\z@skip
  \noindent $^{\scriptstyle\hbox{#1}}$\hskip 4pt%
    \footstrut\futurelet\next\fo@t%
}

\def\footnoterule{\kern-3\p@ \hrule height \z@ \kern 3\p@}

\skip\footins=19.5pt plus 12pt minus 1pt
\count\footins=1000
\dimen\footins=\maxdimen

\def\note#1#2{%
  \let\@sf=\empty \ifhmode\edef\@sf{\spacefactor\the\spacefactor}\/\fi
  #1\insert\footins\bgroup
    \eightpoint\b@ls{10pt}\rm
    \interlinepenalty\interfootnotelinepenalty
    \splitmaxdepth\dp\strutbox \floatingpenalty\@MM
    \leftskip\z@skip \rightskip\z@skip \spaceskip\z@skip \xspaceskip\z@skip
    \noindent\footstrut #1$\,$#2\strut\par
  \egroup
  \@sf\relax}


\def\landscape{%
  \global\TEMPDIMEN=\PageWidth
  \global\PageWidth=\PageHeight
  \global\PageHeight=\TEMPDIMEN
  \global\let\landscape=\relax
  \onecolumn
  \message{(landscape)}%
  \raggedbottom
}


\output{%
  \ifLeftCOL
    \global\setbox\LeftBOX=\vbox to \ZoneBSize{\box255\unvbox\ZoneBBOX
      \ifvoid\footins\else
        \vskip\skip\footins\unvbox\footins\fi
    }%
    \global\LeftCOLfalse
    \MakeRightCol
  \else
    \setbox\RightBOX=\vbox to \ZoneBSize{\box255\unvbox\ZoneBBOX
      \ifvoid\footins\else
        \vskip\skip\footins\unvbox\footins\fi
    }%
    \setbox\MidBOX=\hbox{\box\LeftBOX\hskip\ColumnGap\box\RightBOX}%
    \setbox\PageBOX=\vbox to \PageHeight{%
      \UnloadZoneA\box\MidBOX\UnloadZoneC}%
    \shipout\vbox{\PageHead\vbox to \PageHeight{\box\PageBOX\vss}\PageFoot}%
    \advancepageno
    \ifplate@page
      \shipout\vbox{%
        \sp@pagetrue
        \def\sp@type{plate}%
        \global\plate@pagefalse
        \PageHead\vbox to \PageHeight{\unvbox\plt@box\vfil}\PageFoot%
      }%
      \message{[plate]}%
      \advancepageno
    \fi
    \global\LeftCOLtrue
    \CleanStack
    \MakePage
  \fi
}


\Warn{\start@mess}

\newif\ifCUPmtplainloaded 
\ifprod@font
  \global\CUPmtplainloadedtrue
\fi

\def\mnmacrosloaded{} 

\catcode `\@=12 



\fi
\newif\ifAMStwofonts
\AMStwofontstrue

\ifCUPmtplainloaded \else
  \NewTextAlphabet{textbfit} {cmbxti10} {}
  \NewTextAlphabet{textbfss} {cmssbx10} {}
  \NewMathAlphabet{mathbfit} {cmbxti10} {} 
  \NewMathAlphabet{mathbfss} {cmssbx10} {} 
  \ifAMStwofonts
    \NewSymbolFont{upmath} {eurm10}
    \NewSymbolFont{AMSa} {msam10}
    \NewMathSymbol{\upi}     {0}{upmath}{19}
    \NewMathSymbol{\umu}     {0}{upmath}{16}
    \NewMathSymbol{\upartial}{0}{upmath}{40}
    \NewMathSymbol{\leqslant}{3}{AMSa}{36}
    \NewMathSymbol{\geqslant}{3}{AMSa}{3E}

    \let\leq=\leqslant 
    \let\geq=\geqslant 
  \else
    \def\umu{\mu}
    \def\upi{\pi}
    \def\upartial{\partial}
  \fi
\fi


\pageoffset{-2.5pc}{0pc}

\loadboldmathnames



\pagerange{000--000}    
\pubyear{0000}
\volume{000}

\begintopmatter  

\title{Collapse of Primordial Clouds II. The Role of Dark Matter}

\author{Sandra  R. Oliveira, Oswaldo D. Miranda, Jos\'e C. N. de Araujo
        \hfill\break and Reuven Opher}

\affiliation{Instituto Astron\^omico e Geof\'{\i}sico -- Universidade de
S\~ao Paulo \hfill\break
Av. Miguel St\'efano 4200, S\~ao Paulo, 04301-904, SP, Brazil}

\shortauthor{S.R. Oliveira et al.}
\shorttitle{Collapse of Primordial Clouds II. The Role of Dark Matter}

\abstract{

In this article we extend the study performed in our previous article
on the collapse of primordial objects. We here analyze the behavior of the 
physical parameters for clouds ranging from $10^7M_\odot$ to 
$10^{15}M_\odot$. We studied the dynamical
evolution of these clouds in two ways: purely baryonic clouds and clouds
with non-baryonic dark matter included. We start the calculations at the
beginning of the 
recombination era, following the evolution of the structure until the
collapse (that we defined as the time when the density contrast of the
baryonic matter is greater than $10^4$). We analyze the behavior of the
several physical parameters of the clouds (as, e.g., the density contrast
and the velocities of the baryonic matter and the dark matter) as
a function of time and radial position in the cloud. 
In this study all physical processes that are relevant to the dynamical 
evolution of the primordial clouds, as for example photon-drag (due to the 
cosmic background radiation), hydrogen molecular production, besides the 
expansion of the Universe, are included in the calculations. 
In particular we find that the clouds, with dark matter, collapse at  
higher redshift when we compare the results with the purely baryonic models. 
As a general result we find that the distribution of the non-baryonic dark 
matter is more concentrated than the baryonic one. It is important to stress
that we do not take into account the putative virialization of the 
non-baryonic dark matter, we just follow the time and spatial evolution 
of the cloud solving its hydrodynamical equations. We studied also the 
role of the cooling-heating processes in the purely baryonic clouds.}

\keywords{Cosmology: theory -- dark matter -- early Universe.}

\maketitle

\section{Introduction}

The models of structure formation consider that all the structures of the
Universe, such as galaxies and clusters of galaxies, were formed from  
the growth of small primordial density fluctuations present, initially,
in an homogeneous and isotropic Universe, and that by gravitational
instabilities resulted in the structures that we observe today. 
We are going to see below that we just start the study of formation and
evolution of primordial clouds from density perturbations present at 
the beginning of the recombination era.

The first galaxy formation models that included the effects of gas
dissipation were studied by Larson (e.g., 1969, 1974, 1975, 1976).
Recently, some authors used hydrodynamics codes including the gas
dynamics (e.g., de Araujo 1990, de Araujo \& Opher 1988, 1989, 1991, Haiman
{\it et al} 1995, Thoul \& Weinberg 1995, Tegmark {\it et al} 1997, Oliveira 
{\it et al} 1998, hereafter paper I, among others) in order to follow the 
behavior of the baryonic matter and to study the influence of several physical 
processes on the evolution of the primordial clouds.

Gravity is generally believed to be the dominant force in the processes of 
structure formation and the dynamical studies of these systems, based on
observational data, reveal that the constituents of the Universe include not
only luminous matter, but also a far greater amount of dark matter which
influence the luminous matter through its gravitational field. For systems
such as galaxies and clusters, hydrodynamical effects are important in the
process of their formation and thus, hydrodynamics codes which include dark
matter are then the most accurate manner to follow the cosmological evolution
of the primordial clouds (see, e.g., Cen {\it et al} 1990, Cen 1992, 
Cen {\it et al} 1993).

On the other hand, there are several non-gravitational processes that are
important to the evolution of primordial clouds at high redshift, as the
photon-drag due to the cosmic background radiation, the recombination
processes and the hydrogen molecular formation. These physical processes are
important during and after the recombination era. In general, the models that
use hydrodynamical numerical codes, found elsewhere, do not initiate the
calculations at the beginning of the recombination era, when the perturbations
of the baryonic matter can effectively begin to be significantly amplified,
whether they are Jeans unstable, and do not include the above mentioned
physical processes.

In the present article we used the set of hydrodynamic equations of the paper I
to analyze the behavior of the
physical parameters using first only purely baryonic clouds.
We then compare the evolution of the physical parameters including
non-baryonic dark matter. We start all the calculations present here
at the beginning of the recombination era and take into account
all the relevant physical processes present during and after the 
recombination era, besides the expansion of the Universe.

In \S 2 we describe the basic equations, in \S 3 we discuss our results and 
finally in \S 4 we present our conclusions. 

\section{Equations}

We consider spherically symmetric density perturbations which
produces clouds of baryonic matter and dark matter with densities greater
than the density of the Universe. We consider the baryonic matter and
dark matter as two fluids coupled by gravity.

The equations for a purely baryonic matter are described in greater detail in
paper I, and so, here we only summarize them, namely:
$$
{\partial\rho\over \partial t}+{1\over r^2}{\partial\over\partial r}
(r^2\rho v)=0\eqno\stepeq
$$

$$
{D\vec v\over Dt}=-{1\over\rho}\nabla P-\nabla\phi-{\sigma_{_T}
bT_r^4x\over m_pc}\biggl[\vec v-{\dot R(t)\over R(t)}\vec r\biggl]
\eqno\stepeq
$$

$$
{DE\over Dt}={P\over\rho^2}{D\rho\over Dt}-L\eqno\stepeq
$$

$$
{Dx\over Dt}=C\biggl\{\beta e^{-{(B_1-B_2)\over kT_r}}(1-x)-
{a\rho x^2\over m_p}\biggl\}+I\eqno\stepeq
$$

As in paper I, the Eqs.(1)$-$(4) are, respectively, the equation of
conservation of mass, conservation of momentum, energy and ionization
balance. Where 
$\rho$ is the mass density of the cloud, 
$r$ is the radial coordinate, 
$v$ is the velocity. The velocity is written as
$\vec v=\vec v_n+\vec v_{_H}$, where 
$\vec v_n$ is the peculiar velocity of the cloud and 
$\vec v_{_H}=H \vec r$ (where $H\equiv\dot R(t)/R(t)$ is the Hubble parameter,
$R(t)$ is the scale factor and $\dot R(t)$ its time derivative).
The pressure of the cloud is $P=kN\rho T_m(1+x)$ (where $k$ is the Boltzmann
constant, $N$ is the Avogadro's number, $T_m$ is the temperature of the matter,
and $x$ is the degree of ionization), $\sigma_{_T}$ is the Thomson cross
section, $b={4\sigma_{_{SB}}/c}$ (where $\sigma_{_{SB}}$ is the
Stefan-Boltzmann constant), $T_r$ is the temperature of the radiation,
$m_p$ is the mass of the proton, $c$ is the velocity of light.

The gravitational potential is $\nabla\phi={GM(r)/r^2}$, with
$G$ the gravitational constant. The term $M(r)$ is the total mass (i.e. the 
baryonic and non-baryonic dark matter) contained inside the radius $r$ of
the cloud. We here take the total amount of baryonic matter and dark 
matter constant during the evolution of the clouds.

The cooling-heating processes are included in the cooling function $L$. This
term is the summation of four mechanisms; 
$L=L_{_R}+L_{_C}+L_{_{H_2}}+L_\alpha$. Where,
$L_{_R}$ is the cooling due to recombination (see, e.g., Schwarz {\it et al}
1972), $L_{_C}$ is the Compton cooling (see, e.g., Peebles 1968), 
$L_{_{H_2}}$ is the cooling by molecular hydrogen (see, e.g., Lepp \& Shull
1983) and $L_\alpha$ is the Lyman-$\alpha$ cooling (see, e.g., Calberg
1981). The temperatures in our calculations are never greater than
$\sim 10,000 K$ and thus, the cooling-heating mechanisms considered here
are the dominant mechanisms for the thermal history of the baryonic matter
contained inside the clouds.

In order to consider the $DM$ (non-baryonic dark matter) models, we 
included the following equations:

$$
{\partial\rho_{_{DM}}\over \partial t}+{1\over r^2}{\partial\over\partial r}
(r^2\rho_{_{DM}} v)=0\eqno\stepeq
$$

$$
{D\vec v_{_{DM}}\over Dt}=-\nabla\phi\eqno\stepeq
$$

$$
\vec v_{_{DM}}=\vec v_{n_{_{DM}}}+\vec v_{_H}\eqno\stepeq
$$

$$
\nabla\phi={G M(r)\over r^2}\eqno\stepeq
$$
the various terms appearing in the Eqs.(5)-(8) are analogous
to those found in the hydrodynamical equations to the baryonic matter.

We consider that the primordial clouds evolve within a medium of density
$\rho_u$ (the density of the Universe) and so we take into account
this fact in our study; see paper I to the hydrodynamical equations 
of the Universe.

The initial perturbation is taken to be in the form of a power law spectrum
as considered, for example, by Gott \& Rees (1975) 

$$
\delta_i={\delta\rho\over\rho}=\bigl({M\over M_o}\bigl)^{-{1\over2}-
{n\over6}}(1+z_{rec})^{-1}\eqno\stepeq
$$ 
with $M$ the mass of the cloud, $M_o$ the reference mass, $n$ the 
spectral index and $z_{rec}$ the redshift at the beginning of the 
recombination era. It is important to stress that $M$ is the total mass 
of the clouds, i.e., $M=M_{_B}+M_{_{DM}}$ (where $``B"$ refers to the 
baryonic matter and the $``DM"$  refers to the non-baryonic dark matter). 
The normalization of the spectrum of perturbations given above
was made to agree with the COBE data (see e.g. de Araujo \& Opher 1994). 

We consider that initially the density profile of the cloud is ``top hat" 
like, and we begin the calculations when the radiation temperature is 
$\simeq 4,000 K$ ($z\simeq 1500$) where the ionization degree begins to be
significantly lower than one. We use $\Omega_B=0.1$ (the baryonic density
parameter) and different values for $\Omega_{DM}$ (the non-baryonic dark
matter density parameter), as well as different values for the Hubble
constant, $h$ (the Hubble constant in units of $100km\ s^{-1}
Mpc^{-1}$). We consider that the clouds initially expand with a field
velocity given by the Hubble flow.

We follow the evolution of the cloud until its collapse, that we consider as
the time, after the recombination era, when the density contrast of the
cloud is greater than $10^4$. We used in our calculations 1,200
spherical shells for the baryonic matter and 1,200 shells for the dark matter.
Models with 6,000 shells for both components showed no considerable
difference in the results. The complete set of partial differential equations 
has been written in the form of difference scheme by one of us (Miranda 1998)
according to the basic method described by Richtmyer and Morton (1967).

\section{Calculations and Discussion}

The scale of masses studied ranges from $10^7M_\odot$ to $10^{15}M_\odot$. 
We show, in particular, three points in the cloud; the first shell, the 
middle shell of the scheme, and the last shell as a function of time. 
In order to analyze the radial behavior, we show five different times for 
the output of the data: 
$t_i$ --  time at the recombination era, 
$t_2=(t_3-t_i)/2$, 
$t_3=(t_f-t_i)/2$,
$t_4=(t_f-t_3)/2$ and $t_f$, where
$t_f$ is the time of the collapse of the core.

Also, as already mentioned in the paper I, due to the
expansion of the Universe, even the clouds that are already Jeans unstable
and eventually collapse undergo, initially, an expansion phase before
collapsing (see paper I for more details).

We divided our paper in two parts. Firstly we studied purely baryonic clouds
and we analyzed the behavior of the variables along the evolution of time, and 
in different points inside the cloud. We also studied the influence of the 
cooling-heating processes. To do this we studied individually the effects of 
all the four cooling-heating processes and the photon drag (Tables 2$-$4). 
The results are described in subsection 3.1.

Secondly, we included $DM$ and analyzed the behavior of the same variables in 
the same way, except the analysis of the cooling-heating processes. 
The results of baryonic plus $DM$ model are described in subsection 3.2.

\subsection{Purely baryonic clouds}

We now show the results of the purely baryonic models. We ran our models 
for masses ranging from $10^7M_\odot$ to $10^{15}M_\odot$. Here we ran 
the models with $M_o=4\times10^{17}M_\odot$, $h=1$, $\Omega=0.1$ and $n=-1$. 
The normalization of the spectrum used here is consistent with 
$\sigma_8$ (see, e.g., de Araujo \& Opher 1994 for a discussion).
We show these results in the Figs.1-6. It is worth mentioning that a
spectrum such as $\delta\propto M^{-1/3}$ is consistent with observations for
$1<r/(h^{-1}Mpc)<100$ (see for example Kashlinsky \& Jones 1991).

Let us first have a look at the evolution of the density contrast, in the
Fig.1, and also at the evolution of the density, in the Fig.2.
We see that masses within the interval $10^7M_\odot$ to $10^{10}M_\odot$
have a partial collapse, i.e. our collapse criterion is satisfied only
to the internal shells. In particular, for the masses $10^{7}M_\odot$ to
$10^{9}M_\odot$ the density contrast has a slightly oscillation in the 
external shells and, these shells do not reach a density contrast of
$10^4$.

We also analyzed the morphological behavior of the density
profile (Fig. 3), in order to see how it is modified as a
function of time. For the masses ranging from $10^7M_\odot$ to
$10^{10}M_\odot$, the collapse starts in the internal shells, followed by
the outermost shells, as discussed above. For the masses greater than 
$10^{13}M_\odot$, there is no collapse (at least for the normalization 
of the spectrum of perturbations studied here). Along the radius the initial 
``top hat" profile changes rapidly with time for masses smaller than 
$10^{10}M_\odot$. This is related to the fact that masses lower than 
$10^{10}M_\odot$ have a pressure gradient that modify the initial density 
profile. This shows that to assume that the density profile
is ``top hat" throughout the cloud evolution is not a good assumption,
as considered, e.g., by de Araujo \& Opher (1989, 1991, 1994).
We are going to see later on that under certain circumstances the
use of a ``top-hat'' density profile throughout the calculations, as assumed 
by de Araujo \& Opher, can provide us interesting informations.

Let us now have a look at the temperature, pressure and molecular density 
formation (Fig.4). These variables fall with time for all masses because of
the cooling and the initial expansion of the clouds. But when the 
cloud collapses, the temperature increases greatly as expected. An analysis 
of the molecular density versus temperature is shown in Fig.5. It shows 
that the molecular density increases when the temperature is within the 
interval $100K<T<500K$. 

As already explained in paper I, the heat conduction and the Bremsstrahlung
are not taken into account. It is worth stressing that in our study
we do not take, as initial condition, a virialized object as considered by
some authors. Our calculations start at the recombination era when the
temperature of the baryonic matter inside the perturbation has the same
value as that of the radiation. For clouds that collapse at high redshifts,
in general, the principal mechanism that acts in the evolution and collapse
of the perturbations is the Compton heating-cooling (see results presented
in Tables 5-8 in paper I). Thus, during the expansion phase of the
perturbations, when the Compton heating-cooling is efficient, the
temperature of the baryonic matter inside the perturbation is almost the
same temperature of the radiation.

At the redshift collapse of the clouds, the temperature of the matter
increases by a factor 4 or 5 in relation to its correspondent value at the
turn-around. Thus, depending on the power spectrum and on the mass of the
cloud, the shocks and the collapse itself are not able to rise the
temperature of the cloud to $10^4K$ (see, e.g., Table 5 and the discussion
in the Paper I).

Certainly, these results are dependent on the particular way in which we
define the collapse of our models, that is, we consider that an object
collapsed when the density contrast is greater than $10^4$. The results
also depend on the thermal history of the clouds (that depends on the
physical processes here considered). As a result no strong shocks occur
that could eventually heat the clouds to the virial temperature. If we
follow the calculations to higher values of density than the established
by our collapse criterion, the temperature will rise above the values here
obtained. It is possible that to density contrasts higher than $10^5$
strong shocks occur and they would probably heat the gas to values 
of temperature $>10^4$K, as a result other physical
processes, as e.g. Bremsstrahlung, would be important to the evolution of
the clouds.

In Fig.6 we show the evolution, with the time, of the radii of the 
shells chosen to show the evolution of the clouds. It is possible to see
the instant when the turn-around occurs for these shells. 

In Table 1 we present a comparison between our results and the results
by de Araujo (1990) and de Araujo \& Opher (1994) for purely baryonic clouds. 
In this table we also present calculations using the normalization
consistent with $\sigma_8 =1$ (see, e.g., de Araujo \& Opher 1994),
thereby we have to use $M_o=4\times 10^{17}M_\odot$ in Eq.(9).
In fact this normalization is more realistic since structures as large
as $M=10^{13}M_\odot$ may be formed. 

\begintable{1}
\caption{{\bf Table 1.} The redshifts of collapse}
\halign{#\hfil & \qquad \hfil#\hfil\qquad & 
         \hfil#\hfil\qquad & \hfil#\hfil\qquad &
         \hfil#\hfil\qquad & \hfil#\hfil\cr

& \multispan2 $M_o=4\times10^{17}M_\odot$  
& & \multispan2 $M_o=10^{15}M_\odot$ \cr

$M/M_\odot$ & $z_c^*$ & $z_c^{A^*}$ & & $z_c$ & $z_c^A$ \cr
\noalign{\vskip 10pt}
$10^7$    & 437 & --  & & 102 & 108 \cr
$10^8$    & 238 & --  & & 49  & 49  \cr
$10^9$    & 153 & --  & & 20  & 21  \cr
$10^{10}$ & 77  & 75  & & 6.5 & 7   \cr
$10^{11}$ & 34  & 34  & & NC  & --  \cr
$10^{12}$ & 12  & 13  & & NC  & --  \cr
$10^{13}$ & 2.1 & 3.6 & & NC  & --  \cr
}
\tabletext{We compare here the redshifts of collapse of the present 
article ($z_c^*$), with those by de Araujo (1990) ($z_c^A$) and de Araujo \& 
Opher (1994) ($z_c^{A^*}$). We use $n=-1$, $h=1.0$ and $\Omega_B=0.1$. We
studied two normalizations, namely, $M_o=10^{15}M_\odot$ and 
$M_o=4\times10^{17}M_\odot$. NC stands for ``no collapse".}
\endtable

We note that all masses have collapse redshifts similar to those obtained 
by de Araujo and de Araujo \& Opher using both normalizations, this is a 
very interesting result. If one would like only to have an idea of the 
collapse epoch, the use of a ``top hat'' profile throughout  all the cloud 
evolution is a good approximation (see, e.g., de Araujo \& Opher 1994). 

As mentioned in paper I the clouds stop expanding when
$\bigl({\delta\bar\rho/\rho}\bigr)>4.6$, therefore above the linear 
regime. As seen in paper I, we also obtain in the present calculations that 
the clouds stop expanding when $\bigl({\delta\bar\rho/\rho}\bigr)>4.6$. This
very fact explains why, even in the non-linear regime, some clouds studied
here neither collapses nor stops expanding.

It is worth stressing that the collapse redshift calculated here gives us only 
an idea of the epoch in which the collapse occurs, due to the fact that the 
formation of a galaxy, for example, is a process much more complex.
In fact, what we do, it is a study of formation of a proto-object.
The present study is of interest due to the fact that we show how a density
perturbation could give rise to a proto-object and also we show which
physical processes are important during and after the recombination era.

We now study the influence of the cooling-heating processes on the evolution of 
clouds taking, as representative examples, clouds with the masses $10^7M_\odot$,
$10^9M_\odot$ and $10^{11}M_\odot$ (Tables 2$-$4). To do this, we disregard, 
one at a time, all of the four cooling-heating processes and also the photon 
drag. The results are consistent with paper I, i.e., these very physical 
processes are important on the evolution of the clouds determining, for example,
the collapse epoch and the thermal history of the clouds.

\begintable*{2}
\caption{{\bf Table 2.} Values of the variables for $M=10^7M_\odot$ for 
purely baryonic clouds. We used $h=1.0$, $\Omega_B=0.1$, 
$M_o=10^{15}M_\odot$ and $n=-1$.}
\halign{#\hfil & \quad \hfil#\hfil\quad & 
         \hfil#\hfil\quad & \hfil#\hfil\quad &
         \hfil#\hfil\quad & \hfil#\hfil\quad &
         \hfil#\hfil \cr

         & with all  & without & without  & without & without & without \cr
Variable & processes &$L_{_R}$ & $L_{_C}$ & $L_{_{H_2}}$ & $L_\alpha$
& photon drag \cr
\noalign{\vskip 3pt\hrule\vskip 3pt}
$z_c$ & 103 & 103 & 105 & 103 & 103 & 132 \cr
$n_{_{H_2}}(cm^{-3})$ & $10^2$ & $10^2$ & $1$ & -- & $10^2$ & $300$ \cr
$T (K)$ & $4\times10^2$ & $4\times10^2$ & $9\times10^3$ & $8\times10^3$
& $4\times10^2$ & $5\times10^2$ \cr
$\rho (gcm^{-3})$ & $3\times10^{-19}$ & $3\times10^{-19}$ & $3\times10^{-19}$
& $3\times10^{-19}$ & $3\times10^{-19}$ & $5\times10^{-19}$ \cr
\noalign{\vskip 3pt\hrule\vskip 10pt}
}
\endtable

\begintable*{3}
\caption{{\bf Table 3.} Values of the variables for $M=10^9M_\odot$ for 
purely baryonic clouds. We used $h=1.0$, $\Omega_B=0.1$, 
$M_o=10^{15}M_\odot$ and $n=-1$.}
\halign{#\hfil & \quad \hfil#\hfil\quad & 
         \hfil#\hfil\quad & \hfil#\hfil\quad &
         \hfil#\hfil\quad & \hfil#\hfil\quad &
         \hfil#\hfil \cr

          & with all  & without & without & without & without & without \cr
Variable  & processes &$L_{_R}$ &$L_{_C}$ &$L_{_{H_2}}$ & $L_\alpha$ 
& photon drag \cr
\noalign{\vskip 3pt\hrule\vskip 3pt}
$z_c$                  & 20 & 20 & 228       & 20  & 20 & 28 \cr
$n_{_{H_2}}(cm^{-3})$  & 3  & 3  & $10^{-1}$ & --  & 3  & 1  \cr
$T (K)$ & $5\times10^2$ & $5\times10^2$ & $4\times10^2$
& $1\times 10^4$ & $6\times10^2$ & $5\times10^2$ \cr
$\rho (gcm^{-3})$ & $3\times10^{-21}$ & $3\times10^{-21}$ & $3\times10^{-21}$
& $3\times10^{-21}$ & $3\times10^{-21}$ & $5\times10^{-21}$ \cr
\noalign{\vskip 3pt\hrule\vskip 10pt}
}
\endtable

\begintable*{4}
\caption{{\bf Table 4.} Values of the variables for $M=10^{11}M_\odot$ for 
purely baryonic clouds. We used $h=1.0$, $\Omega_B=0.1$, 
$M_o=10^{15}M_\odot$ and $n=-1$.}
\halign{#\hfil & \quad \hfil#\hfil\quad & 
         \hfil#\hfil\quad & \hfil#\hfil\quad &
         \hfil#\hfil\quad & \hfil#\hfil\quad &
         \hfil#\hfil \cr

         & with all  & without  & without  & without & without & without \cr
Variable & processes & $L_{_R}$ & $L_{_C}$ &$L_{_{H_2}}$ & $L_\alpha$ 
& photon drag \cr
\noalign{\vskip 3pt\hrule\vskip 3pt}   
$z_c$ & NC & NC & 0.14 & NC & NC & 2.6 \cr
$n_{_{H_2}}(cm^{-3})$  & $1\times 10^{-9}$ & $3\times10^{-8}$
& $1\times 10^{-6}$ & -- & $3\times10^{-8}$ & $3\times10^{-5}$ \cr
$T (K)$ & $10$ & $50$ & $10$ & $10$ & $50$ & $4\times10^2$ \cr
$\rho (gcm^{-3})$ & $3\times10^{-28}$ & $1\times 10^{-26}$ 
& $1\times 10^{-25}$ & $1\times 10^{-26}$ & $1\times 10^{-26}$
& $3\times10^{-24}$ \cr
\noalign{\vskip 3pt\hrule\vskip 10pt}
}
\endtable

The analysis of the cooling mechanisms show that when we disregard the
Compton cooling-heating, the collapses occur earlier.
This is due to the fact that the Compton mechanism heats the cloud when the 
temperature of the matter is lower than the temperature of the radiation,
in this way during the recombination era the Compton cooling-heating
maintains the temperature of the cloud near to the temperature of the
cosmic background radiation. Due to the expansion of the Universe
the clouds firstly expand and whether the Compton heating was not present
the clouds would expand almost adiabatically, as a result their temperatures
would get smaller, and so their pressures, thereby the clouds would stop
expanding earlier.

When we disregard the photon drag the clouds also collapse earlier. 
As we show in paper I the photon drag acts distinctly on the smaller
masses ($< 10^6M_\odot$) and on the greater masses. Here it acts against 
the collapse, as a brake, but its action is not so strong like it was 
for the smaller masses. 

The cooling by molecular hydrogen does not influence the collapse redshift,
its influence is on the thermal history of the cloud. The other physical
processes here studied have a behavior similar to that described in paper I.

We studied also the behavior of the collapse of the clouds when we change 
$h$ and $\Omega$. The results are in Tables 5 and 6. 
As a general conclusion it can be mentioned that the collapse redshift, 
the maximum mass of the objects formed, and also the thermal history
of the clouds depend strongly on $h$ and $\Omega$. When we change, for 
example, $\Omega$ from  0.1 to 0.2 we have collapses occurring earlier. 
On the other hand, when we change $h$ from 1 to 0.5 we have the collapses
occurring latter, as expected. 

\begintable{5}
\caption{{\bf Table 5.} Values of the variables for $h=1.0$, $\Omega_B=0.2$, 
$M_o=10^{15}M_\odot$ and $n=-1$ for purely baryonic clouds.}
\halign{#\hfil & \qquad \hfil#\hfil\qquad & 
         \hfil#\hfil\qquad & \hfil#\hfil\qquad &
         \hfil#\hfil \cr

$M/M_\odot$ & $z_c$ & $n_{_{H_2}}(cm^{-3})$ & $T (K)$ & $\rho (gcm^{-3})$
\cr
\noalign{\vskip 10pt}
$10^7$    & 110 & 20 & $4\times10^2$ & $3\times 10^{-20}$ \cr
$10^9$    & 24  & 1  & $6\times10^2$ & $1\times 10^{-20}$ \cr
$10^{11}$ & 2   & $2\times 10^{-5}$ & $3\times 10^2$
& $3\times 10^{-24}$ \cr
\noalign{\vskip 10pt}
}
\endtable

\begintable{6}
\caption{{\bf Table 6.} Values of the variables for $h=0.5$, $\Omega_B=0.1$, 
$M_o=10^{15}M_\odot$ and $n=-1$ for purely baryonic clouds.}
\halign{#\hfil & \qquad \hfil#\hfil\qquad & 
         \hfil#\hfil\qquad & \hfil#\hfil\qquad &
         \hfil#\hfil \cr

$M/M_\odot$ & $z_c$ & $n_{_{H_2}}(cm^{-3})$ & $T (K)$ & $\rho (gcm^{-3})$
\cr
\noalign{\vskip 10pt}
$10^7$    & 93 & 300 & $5\times 10^2$ & $5\times 10^{-19}$ \cr
$10^9$    & 18 & 500 & $6\times 10^2$ & $3\times 10^{-22}$ \cr
$10^{11}$ & NC & $10^{-10}$ & 10 & $3\times 10^{-29}$ \cr
}
\endtable

\subsection{Baryonic plus DM clouds}

We now discuss the results of our models including the presence of the
dark matter. 
We ran our models for masses ranging from $10^7M_\odot$ to $10^{15}M_\odot$. 
Here we ran initially all models with $M_o=4\times10^{17}M_\odot$, $h=0.5$, 
$\Omega_{_B}=0.1$, $\Omega_{_{DM}}=0.5$ and $n=-1$, and we show these results 
in the Figs.7$-$13. In order to compare we also ran models with
different combinations of $h$ and $\Omega_{_{DM}}$, as well as, a different 
normalization of the spectrum of density perturbations.

The behavior of the density is shown in Fig.7 (time evolution) 
and Fig.8 (spatial evolution). 
For the normalization of the spectrum of perturbations used here, 
the maximum mass that collapses is $10^{14}M_\odot$. 

It is worth mentioning that the explanation of why there is no collapse for
some clouds, as could be expected, is due to the fact, as already mentioned,
that it is necessary to have $\bigl({\delta\bar\rho/\rho}\bigr)>4.6$ to the
cloud stops expanding and starts its collapse.

We see, from Fig.7, rigth side, that all masses have partial collapse for 
the baryonic component, the outermost shells expand with the Universe. 
For the $DM$ component (left side of the Fig.7), this partial collapse occurs
only for the cloud of $10^{14}M_\odot$. 

In Fig.8 we see that the top hat profile is destroyed as time goes on,
particularly to the baryonic matter. 
In the $DM$ component (left side of Fig.8) 
the top hat profile is maintained almost all the time during its evolution,
but it is destroyed rapidly when the collapse is more evolved, in particular
for the cases in which we have $M > 10^{12} M_\odot$.  

The DM profile is maintained due to the fact that it is 
pressure free, but
when the gravitational influence of the baryonic matter, that is no pressure
free, begins to be important the top hat profile of the $DM$ component is
destroyed. 

The behavior of the velocity profile is shown in Fig.9 (time evolution)
and Fig.10 (spatial evolution). In the baryonic matter component
(right side of Fig.10), the profile of the velocity begins linear but it
changes rapidly with time, just like in the purely baryonic model. 
The velocity profile of the $DM$ component (left side of Fig.12) is
maintained linear almost all the time of the cloud evolution, only for clouds
of $M > 10^{12} M_\odot$ there is a slightly modification in such
a behavior.

The temperature, pressure and molecular density formation of the baryonic
matter component is shown in Fig.11. As can be seen in Figs.11-12,
the production of the $H_2$ molecules is efficient for all masses, once
reached the temperature, of the baryonic matter, in the range
$100K<T<500K$. Near $z_c$ the $H_2$ production increases, but for the external
shells (long dashed line) it is always low, due to the fact that the
density of the cloud is also low in such shells. 

In Fig.13 we show the behavior of the radii of the shells for the $DM$ 
component  and for the baryonic matter component (left side and right 
side of the figure, respectively). Comparing the two matter distribution
it can be seen that the $DM$ matter component is more concentrated
than the baryonic matter component.

At this point it is worth stressing that in our calculations we
perform a purely hydrodynamical study of the evolution of the clouds,
we do not consider that neither the $DM$ nor the baryonic matter undergo,
for example, virialization. 

Due to the fact that there is no virialization for the $DM$, and that 
it is pressure free, it does not either undergo the photon-drag effect,
and thus the $DM$ gets more concentrated. 

Its well known, and widely accepted, that the spiral galaxies have a halo with 
at least part of it made of non-baryonic dark matter (see, e.g., Persic \& 
Salucci 1990). In this case the $M_B(r)/M_{DM}(r)$ decreases with increasing 
$r$, contrary to the results present here, although we here are very far to 
have a complete model for a galaxy. Virialization and star formation, for 
example, will certainly modify strongly the history of the proto-objets here 
studied. The standard scenario of galaxy formation outlined by, for example,
White \& Rees (1978) explains what could happen with the relative 
distribution of dark and baryonic matter. The dark matter forms a halo
that collapses, and the gas shock heats to the virial temperature
and is pressure supported. The gas pressure then decreases through
radiative cooling and then sinks leaving behind a halo of dark matter.

In elliptical galaxies the distribution of dark matter is not well determined
as in the spiral galaxies. Based on the X-ray observations of hot gas of
ellipticals, probably the best source of information in this case, some
authors argued that $M_{_{BDM}} \sim r$ (where $BDM$ is the baryonic dark
matter), just like in the spiral galaxies. It is worth stressing, however,
that these studies assume that the hot gas is isothermal, but usually the
temperature profile is poorly known (see, e.g., Carr 1994). Whether the
non-baryonic dark matter and the baryonic dark matter have the same
distribution, and whether the ellipticals follow the same $DM$ distribution
followed by the spirals, again there is no accordance between our results
and the distribution of $DM$ inferred for the ellipticals. On the other hand,
we can argue that we are very far to have a complete model for an elliptical
galaxy, to justify the lack of accordance.

From X-ray observations it was shown that the cluster of galaxies
A665, for example, presents a high concentration of (baryonic) dark 
matter at its center (see, e.g., Hughes \& Tanaka 1992). Similar high 
concentration of dark matter at the center of clusters was derived
for the Perseus (Eyles 1992) and Coma (Briel, Henry \& Bohringer 1992)
among other clusters. Some authors argued that this dark matter should
be dissipative and therefore baryonic, but it could well be, at least 
part of it, of non-baryonic nature. In this case the  non-baryonic dark
matter to baryonic matter distribution found in our study for structures
with $M > 10^{14} M_\odot$ would be in accordance with the distribution
of $DM$ inferred from the X-ray observations of clusters of galaxies,
whether the non-baryonic dark matter follows a distribution similar
to that followed by the baryonic dark matter.

We also ran models with other combinations of the parameters $n$, $h$, 
$\Omega_{DM}$ and $M_o$ to see how, for example, $z_c$ and the maximum 
mass that collapses are modified. We show these results in Tables 7$-$9.

In Table 7 we show the collapse redshift, $z_c$, for $h = 1$ and 0.5
for the normalization $M_o=4\times10^{17}M_\odot$, as well as for 
the normalization $M_o=10^{15}M_\odot$. For these cases we assume 
a Universe with $\Omega_B = 0.1$ and $\Omega_{DM} = 0.5$.

\begintable{7}
\caption{{\bf Table 7.} Models with baryonic plus dark matter}
\halign{#\hfil & \quad \hfil#\hfil\quad & 
         \hfil#\hfil\quad & \hfil#\hfil\quad &
         \hfil#\hfil \cr

&\multispan2 $M_o=4\times10^{17}M_\odot$
&\multispan2 $M_o=10^{15}M_\odot$ \cr
$M/M_\odot$ & $z_c$  & $z_c$ & $z_c$  & $z_c$  \cr
& $(h=0.5)$ & $(h=1)$ & $(h=0.5)$ & $(h=1)$ \cr
\noalign{\vskip 10pt}
$10^7$    & 515 & 518 & 131 & 132 \cr
$10^8$    & 332 & 335 & 66  & 67  \cr
$10^9$    & 191 & 193 & 31  & 32  \cr
$10^{10}$ & 100 & 101 & 14  & 14  \cr
$10^{11}$ & 49  & 50  & 5.7 & 5.7 \cr
$10^{12}$ & 22  & 23  & 1.6 & 1.6 \cr
$10^{13}$ & 9.4 & 9.5 & NC  & NC  \cr
$10^{14}$ & 2.5 & 2.2 & NC  & NC  \cr
$10^{15}$ & NC  & NC  & NC  & NC  \cr
}
\tabletext{Collapse redshifts of baryonic plus $DM$ models with
$\Omega_{_B}=0.1$, $\Omega_{_{DM}}=0.5$ and $n=-1$.} 
\endtable

\begintable{8}
\caption{{\bf Table 8.} Collapse redshifts of the models}
\halign{#\hfil & \qquad \hfil#\hfil\qquad & 
         \hfil#\hfil\qquad & \hfil#\hfil\qquad &
         \hfil#\hfil \cr

&\multispan2 {only baryonic }
&\multispan2 {baryonic plus $DM$} \cr
$M/M_\odot$ & $z_{c_{_B}}$ & $z_{c_{_{94}}}$ & $z_{c_{_{DM}}}$
& $z_{c_{_{91}}}$ \cr
\noalign{\vskip 10pt}
$10^7$    & 102 & 108 & 132 & 113 \cr
$10^8$    & 49  & 49  & 67  & 56  \cr
$10^9$    & 20  & 21  & 32  & --  \cr
$10^{10}$ & 6.5 & 7   & 14  & --  \cr
$10^{11}$ & NC  & NC  & NC  & NC  \cr
}
\tabletext{Comparison of the collapse redshifts of only baryonic 
models ($z_{c_{_B}}$), baryonic plus $DM$ models 
($z_{c_{_{DM}}}$) and de Araujo \& Opher's (1991, 1994) 
($z_{c_{_{91}}}$ and $z_{c_{_{94}}}$) models with
$M_o=10^{15}M_\odot$, $n=-1$, $h=1$, $\Omega_{_B}=0.1$,
$\Omega_{_{DM}}=0.5$.}
\endtable

\begintable{9}
\caption{{\bf Table 9.} Comparison of the collapse redshifts of baryonic
plus $DM$ models with $M_o=4\times10^{17}M_\odot$, $h=0.5$ and
$\Omega_{_B}=0.1$.} 
\halign{#\hfil & \quad \hfil#\hfil\quad & 
         \hfil#\hfil\quad & \hfil#\hfil \cr

$M/M_\odot$ & $n=-1$  & $n=-1$ & $n=+1$  \cr
& $(\Omega_{_{DM}}=0.5)$ & $(\Omega_{_{DM}}=0.8)$ & $(\Omega_{_{DM}}=0.5)$
\cr
\noalign{\vskip 10pt}
$10^7$    & 515 & 520 & 1471 \cr
$10^8$    & 332 & 336 & 1435 \cr
$10^9$    & 191 & 194 & 1366 \cr
$10^{10}$ & 100 & 102 & 1235 \cr
$10^{11}$ & 49  & 50  & 997  \cr
$10^{12}$ & 22  & 23  & 635  \cr
$10^{13}$ & 9.4 & 9.9 & 270  \cr
$10^{14}$ & 2.5 & 2.8 & 77   \cr
$10^{15}$ & NC  & NC  & 14   \cr
\noalign{\vskip 10pt}
}
\endtable

For the normalization $M_o=4\times10^{17}M_\odot$ we obtain
no significant differences for $z_c$ when we change $h = 1$ to 0.5.
Also, the maximum mass for both cases is $M_{max}=10^{14}M_\odot$.

For the normalization $M_o=10^{15}M_\odot$ we also obtain
no significant differences for $z_c$ when we change $h = 1$ to 0.5, 
for all clouds. 

Comparing now the results for the two normalizations, we obtain that 
the values of $z_c$ are higher for $M_o=4\times 10^{17}M_\odot$.
These results can be understood due to the fact that the normalization
$M_o=4\times 10^{17}M_\odot$ have a greater initial density contrast when we
compare with $M_o=10^{15}M_\odot$ and thus, for the same values
of $h$ and $\Omega$, the normalization $M_o=4\times 10^{17}M_\odot$ produces 
collapse of the clouds before those with the normalization 
$M_o=10^{15}M_\odot$.

In Table 8 we compare our purely baryonic models and our baryonic plus 
$DM$ models with the models by de Araujo \& Opher (1991, 1994). 
For the purely baryonic cases, and in particular for the range of 
masses studied, there is no significant difference between
the results for $z_c$ present here as compared with de Araujo \& Opher's 
results. For the cases with non-baryonic dark matter we obtain 
values  for $z_c$ larger than that obtained by de Araujo \& Opher.
Comparing now only the results of our studies, we obtain that the 
inclusion of dark matter makes the collapses to occur earlier.

From Table 9 we compare the results of the models when we change 
the slope of the spectrum ($n$) and the $DM$ content ($\Omega_{_{DM}}$). 
Comparing the values of $z_c$ when we use $\Omega_{_{DM}}=0.8$, instead of
$\Omega_{_{DM}}=0.5$, we obtain no significant differences in the results.
Also, the maximum mass in both cases is $M_{max} = 10^{14} M_\odot$.
Comparing now $z_c$, for $\Omega_{_{DM}}=0.5$, with $n=-1$ and $n=+1$
we obtain very different values. The $z_c$ values are 
very high for all masses studied when we adopt $n=+1$; in fact a power 
spectrum with such a spectral index produces excessive power. We have 
for this case collapses of masses of up to $M_{max}=10^{15}M_\odot$.

\section{Conclusions}

In the present investigation, we studied the evolution of density 
perturbations with the masses ranging from $10^7 - 10^{15}M_\odot$,
taking into account a series of physical processes which are present during
and after the recombination era. In order to perform such a study, a
spherical Lagrangian hydrodynamical code was written, which made possible to
follow the spatial and time evolution of the density perturbations with 
purely baryonic and baryonic plus non baryonic dark matter clouds.

Our main conclusions for purely baryonic models are the following:

\beginlist
\item a) Clouds with masses that range from $10^7-10^{10}M_\odot$ have a
partial collapse, for $M_o=10^{15}M_\odot$;
\item b) Clouds having masses $>10^{13} M_\odot$ do not have enough time
to collapse, for $M_o=10^{15}M_\odot$;
\item c) If one changes the perturbation spectrum we have collapse for 
masses up to $10^{13} M_\odot$;
\item d) The Compton cooling-heating plays a key role in the collapse of the 
clouds; 
\item e) the photon-drag delays the collapse and 
\item f) The $H_2$ molecules are, in general, very important to the thermal
history of the cloud.
\endlist

Our main conclusion for the baryonic plus $DM$ models are the following:

\beginlist
\item a) We have earlier collapse redshifts always, as compared with the
purely baryonic models, independent of the combination of the values of $n$, 
$h$, $\Omega_{DM}$ and $M_o$ that we used; 
\item b) In the collapse of the $DM$ the cooling-heating mechanisms and the 
photon drag, which are very important for the collapse of the baryonic 
matter, cannot directly alter the collapse of this component and thus the 
non-baryonic matter is rapidly concentrated in the innermost region of the 
clouds;  
\item c) The observations suggest that in clusters of galaxies the
dark matter is more concentrated than the baryonic matter.
This result is in agreement with our results if the dark matter is 
of non-baryonic origin. 
\endlist

\section*{Acknowledgements}

We ran the models on a HP--Apollo 9000 (purchased by the Brazilian agency 
FAPESP) and on the CRAYs EL98 and J90 (CCE/USP).    

We thank the referee, Dr. David Weinberg, for helpful comments and 
suggestions that greatly improved the present version of our paper.

We thank the Brazilian agencies CAPES (SRO) and CNPq (ODM, JCNA and RO)
for support.

\vskip 1truecm
\noindent{\bf FIGURE CAPTIONS}
\vskip 1truecm 

{\bf Fig.1 --}The evolution of the density contrast ($\delta$) with time
for the mass perturbations in the range $10^7-10^{12}M_\odot$. The solid
line is the internal shell, the short dashed line is the middle shell
and the long dashed line is the external shell, where the time is given
in years. The input parameters of the models are: $M_o=4\times10^{17}
M_\odot$, $h=1.0$, $\Omega_{B}=0.1$ and $n=-1$.

{\bf Fig.2 --}The evolution of the mass density $\rho$ (in $g \ cm^{-3}$)
for the baryonic matter with time. The captions and input parameters for the
curves are the same as the Fig.1.

{\bf Fig.3 --}The mass density $\rho$ versus radius (in $pc$) of the
clouds. The solid line is the initial time ($t_i$), the short dashed line,
long dashed line and very long dashed line are intermediate times
(respectively $t_2$, $t_3$ and $t_4$) and the dashed-pointed line is the
final time ($t_f$). The input parameters are the same as the Fig.1.

{\bf Fig.4 --}The evolution of the temperature (T in $K$), pressure of
the gas (P in $dyn \ cm^{-2}$) and density of molecules ($n_{H_2}$ in
$molecules \ cm^{-3}$). The captions and input parameters for the curves
are the same as the Fig.1.

{\bf Fig.5 --}The evolution of the density of molecules ($n_{H_2}$) versus  
temperature (T). The captions and input parameters for the curves are the
same as the Fig.1.

{\bf Fig.6 --}The evolution of the radii of the purely baryonic clouds
for $10^7-10^{12}M_\odot$ as a function of time. The solid line is the
internal shell, the short dashed line is the middle shell and the long
dashed line is the external shell, where the time is given in
years. The input parameters for these models are the same as the Fig.1.

{\bf Fig.7 --}The evolution of the mass density ($\rho$) with time.
The solid line is the internal shell, the short dashed line is the middle
shell and the long dashed line is the external shell, where the time is
given in years. The input parameters for these models are:
$M_o=4\times10^{17}M_\odot$, $h=0.5$, $\Omega_{B}=0.1$, $\Omega_{DM}=0.5$
and $n=-1$.

{\bf Fig.8 --}The mass density ($\rho$) versus the radii of the clouds.
The solid line is the initial time ($t_i$), the short dashed line,
long dashed line and very long dashed line are intermediate times
(respectively $t_2$, $t_3$ and $t_4$) and the dashed-pointed line is the
final time ($t_f$). The input parameters are the same as the Fig.7.

{\bf Fig.9 --}The evolution of the velocity (in $km\;s^{-1}$) with time.
The captions and input parameters for the curves are the same as the Fig.7.

{\bf Fig.10 --}The velocity profiles to the baryonic and dark matter
components. The captions for the curves are the same as the Fig.8. The input
parameters for the models are: $M_o=4\times10^{17}M_\odot$, $h=0.5$,
$\Omega_{BR}=0.1$, $\Omega_{DM}=0.5$ and $n=-1$.

{\bf Fig.11 --}The evolution of the temperature (T), pressure (P) and 
density of molecules ($n_{H_2}$) with time of the baryonic matter component.
The captions and input parameters for the curves are the same as the
Fig.7. 

{\bf Fig.12 --}The temperature of the gas (T) versus the molecular
density ($n_{H_2}$). The captions and input parameters for the curves are
the same as the Fig.7. 

{\bf Fig.13 --}The evolution of the radius of the baryonic and dark matter
components of the clouds with time. The captions and input parameters for
the curves are the same as the Fig.7.

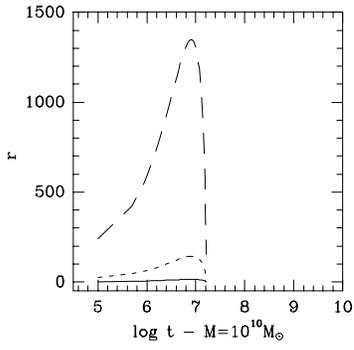
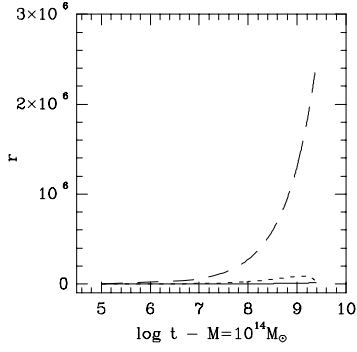
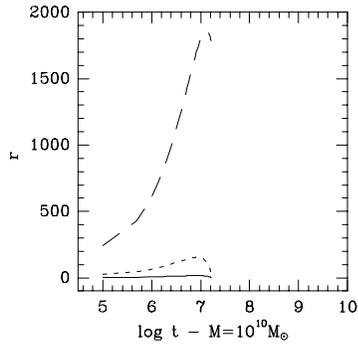
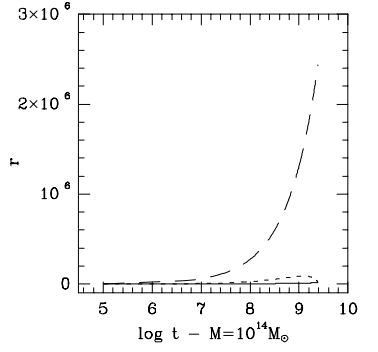
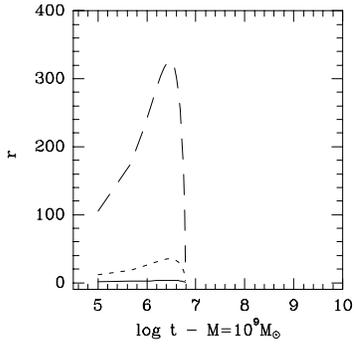
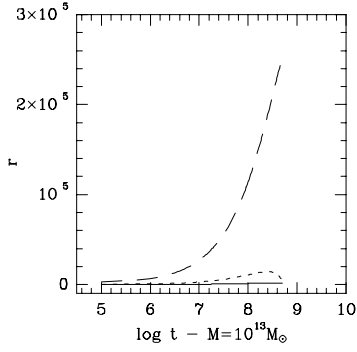
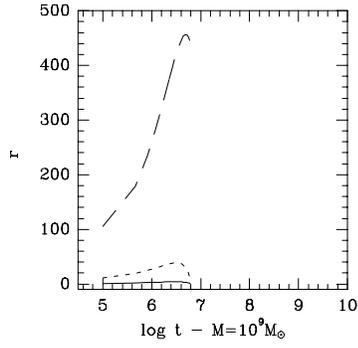
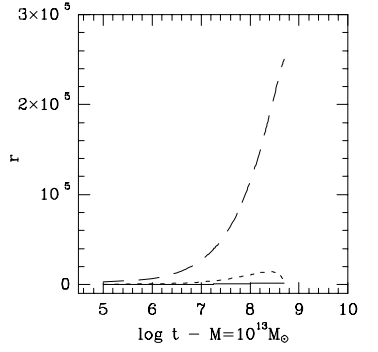
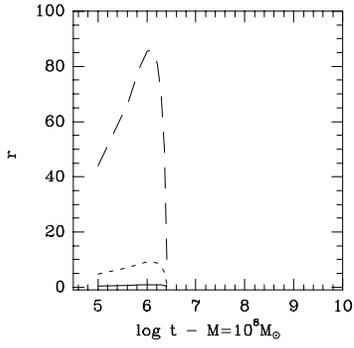
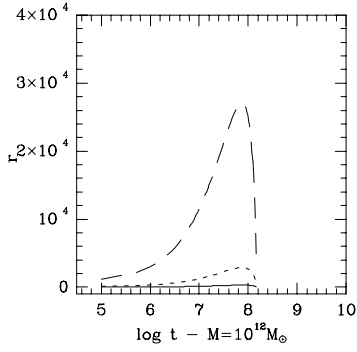
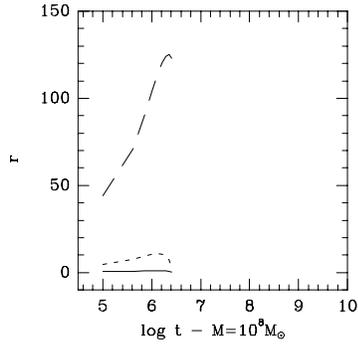
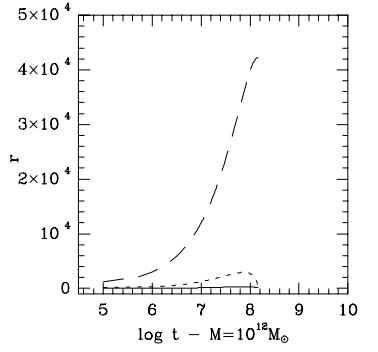
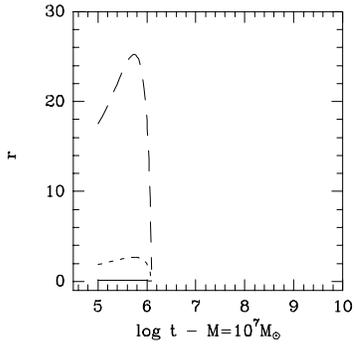
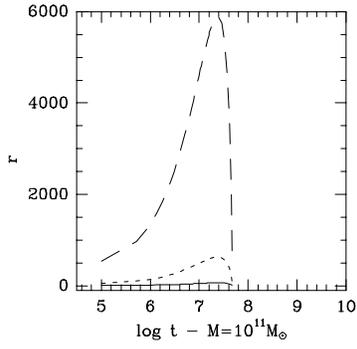
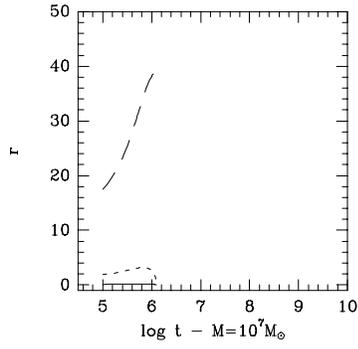
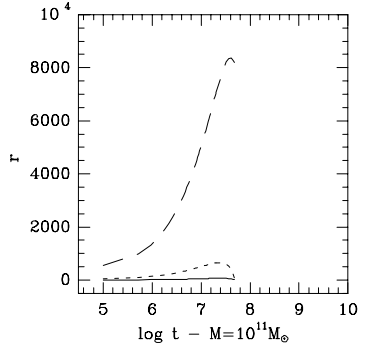
\bye